\documentclass[11pt]{article}

\usepackage[utf8]{inputenc}
\usepackage[T1]{fontenc}
\usepackage{lmodern}
\usepackage{microtype}
\usepackage{amsmath,amssymb}
\usepackage{graphicx}
\usepackage{booktabs}
\usepackage{array}
\usepackage{multirow}
\usepackage{caption}
\usepackage[margin=1in]{geometry}
\usepackage[hidelinks]{hyperref}
\usepackage{url}
\graphicspath{{figures/}}

\title{Quantifying Uncertainty in AI Visibility\\[4pt]
\large A Statistical Framework for Generative Search Measurement}
\author{Ronald Sielinski\\ IQRush\\ \texttt{ron@iqrush.ai}}
\date{}

\begin{document}
\maketitle

\begin{abstract}
AI-powered answer engines are inherently non-deterministic: identical queries submitted at different times can produce different responses and cite different sources. Despite this, current approaches to measuring domain visibility in generative search rely on single-run point estimates of citation share and prevalence, treating them as fixed values. This paper argues that citation visibility metrics should be treated as sample estimators of an underlying response distribution. We conduct an empirical study of citation variability across three generative search platforms---Perplexity Search, OpenAI SearchGPT, and Google Gemini---using repeated sampling across three consumer product topics under two sampling regimes: daily collections over nine days and high-frequency sampling at ten-minute intervals. Citation distributions follow a power-law form and exhibit substantial variability across repeated samples. Bootstrap confidence intervals reveal that many apparent differences between domains fall within the noise floor of the measurement process. Distribution-wide rank stability analysis further demonstrates that citation rankings are unstable across samples, not only among top-ranked domains but throughout the frequently cited domain set. These findings show that single-run visibility metrics provide a misleadingly precise picture of domain performance in generative search. We argue that citation visibility must be reported with uncertainty estimates and provide practical guidance on sample sizes required to achieve interpretable confidence intervals.
\end{abstract}


\medskip
\noindent\textbf{Additional Keywords and Phrases:} AI visibility, answer engine optimization, generative engine optimization.

\medskip

\hypertarget{introduction}{%
\section{Introduction}\label{introduction}}

Generative search engines have transformed how users encounter
information online. Rather than returning a ranked list of hyperlinks,
platforms such as Perplexity Search, OpenAI SearchGPT, and Google Gemini
synthesize a natural-language response to each query and attribute
claims within that response to a set of cited web sources. For brand
managers, digital marketers, and content strategists, these citations
have become a new form of visibility: appearing in a cited source
confers authority and drives traffic in a way that a traditional search
ranking no longer guarantees.

Practitioners have responded by developing measurement frameworks that
report, for example, citation counts, citation share, and citation
prevalence as metrics of AI visibility. These reports typically derive
from a single batch of queries submitted on a single occasion. The
resulting numbers are treated as ground truth: domain A has 12\%
citation share, domain B has 8\%, and A is therefore more visible.
Optimization campaigns are launched, interventions are measured, and
conclusions are drawn---all on the basis of single-run point estimates.

This paper argues that such measurements are fundamentally unreliable,
because the systems they measure are non-deterministic. The same query
submitted twice to the same platform can produce a different cited
source set. This is not a bug or an edge case: it is a design property
of generative AI systems, which use stochastic sampling at generation
time. The practical consequence is that citation visibility metrics are
random variables, not fixed values, and any single measurement of them
carries unquantified uncertainty that may be large enough to invalidate
common analytical conclusions.

\hypertarget{motivating-example}{%
\subsection{Motivating Example}\label{motivating-example}}

Consider a concrete decision a practitioner faces every day. A brand
manager measuring domain visibility on SearchGPT for the topic of
running gear collects a sample of 200 queries and observes that
\emph{tomsguide.com} has a citation share of approximately 9.5\%, while
\emph{runnersworld.com} has a citation share of approximately 6.0\%. The
conclusion seems clear: Tom's Guide is the dominant source for this
topic on this platform, outperforming Runner's World by more than three
percentage points. Resources are allocated accordingly.

The 95\% bootstrap confidence intervals tell a different story. The
interval for \emph{tomsguide.com} spans approximately 5.5\% to 12.5\%;
the interval for \emph{runnersworld.com} spans approximately 4.0\% to
8.0\%. These intervals overlap substantially: the entire
\emph{runnersworld.com} interval falls within the \emph{tomsguide.com}
interval. The apparent difference of 3.5 percentage points is well
within the range of sampling noise. A practitioner cannot conclude from
this sample that \emph{tomsguide.com} outperforms
\emph{runnersworld.com} with any statistical confidence.

This is not a pathological edge case. Across the platforms and topics
studied in this paper, overlapping confidence intervals of this kind are
the norm rather than the exception for domains that appear to differ in
citation share by less than 5-7 percentage points. The bootstrap
confidence interval makes the uncertainty explicit and actionable: it
transforms an apparently decisive ranking into a result that is
statistically indistinguishable from a tie. Citation visibility metrics
must be accompanied by uncertainty estimates, or they are liable to
mislead.

This paper reframes the measurement of domain visibility in generative
search as a statistical estimation problem for stochastic information
systems and demonstrates empirically that reliable visibility
measurement requires repeated sampling and explicit uncertainty
quantification.

\hypertarget{contributions}{%
\subsection{Contributions}\label{contributions}}

This paper makes the following contributions:

\begin{itemize}
\item
  An empirical characterization of citation variability in three
  generative search platforms across repeated sampling, demonstrating
  that stochastic system behavior produces material differences in
  measured domain visibility.
\item
  A formal framework distinguishing system-level stochasticity (the
  answer engine produces different outputs on repeated calls) from
  measurement uncertainty (estimates derived from finite samples are
  imprecise) and showing that both are present and consequential.
\item
  A demonstration that citation count distributions follow a power-law
  form, and that variability differs structurally between the head and
  tail of the distribution---with implications for how frequently and
  infrequently cited domains should be analyzed.
\item
  An application of bootstrap confidence intervals to citation
  visibility metrics, showing that they are both necessary (intervals
  are wide relative to observed differences) and computationally
  tractable.
\item
  A distribution-wide rank stability analysis using weighted Spearman
  rank correlation, demonstrating that citation rank instability extends
  across the full frequently-cited domain set and is not confined to the
  top-ranked domains, establishing that confidence intervals are
  necessary for inference at any rank position in the distribution.
\item
  A methodological control using content checksums to verify that
  observed variability is attributable to the answer engines rather than
  to changes in the content of cited pages.
\end{itemize}

\hypertarget{related-work}{%
\section{Related Work}\label{related-work}}

\hypertarget{citation-quality-in-generative-search}{%
\subsection{Citation Quality in Generative
Search}\label{citation-quality-in-generative-search}}

A growing body of work examines the accuracy and faithfulness of
citations in generative search outputs. Liu et al. establish that
generative systems frequently cite sources that do not support the
attributed claims and introduce metrics for citation recall and
precision \cite{liu2023}. Gao et al. present the ALCE benchmark for automatic
citation evaluation, which operationalizes citation quality as the
degree to which cited passages entail generated claims \cite{gao2023}. Xu et
al. extend this framework with CiteBench, introducing fine-grained
criteria for evaluating citation faithfulness across multiple dimensions
\cite{xu2025}. Li et al. propose AttributionBench, targeting the attribution
verification problem specifically \cite{li2024}.

This literature treats citation accuracy as the central measurement
challenge: given a response and its citations, do the citations support
the claims? Implicit in this framing is that each measurement is a
reliable observation about a well-defined system state. None of these
works address the variability of citation selection across repeated runs
of the same query---the question of whether the set of cited sources
itself is stable. Our work complements this literature by characterizing
the stochastic properties of the citation process rather than the
quality of individual citations.

\hypertarget{visibility-and-optimization-in-generative-search}{%
\subsection{Visibility and Optimization in Generative
Search}\label{visibility-and-optimization-in-generative-search}}

Aggarwal et al. introduce the GEO framework and the GEO-BENCH evaluation
suite, which defines domain visibility metrics and reports the effects
of content interventions on citation rates across answer engines
\cite{aggarwal2024}. GEO is the most directly competitive work to ours: it defines
visibility metrics that align closely with citation share and citation
prevalence as we define them, and it reports statistically significant
improvements from content interventions.

The critical gap is the absence of confidence intervals on both the
baseline and post-intervention measurements. Without uncertainty
quantification, it is impossible to determine whether reported
visibility improvements exceed the noise floor of the measurement
process itself. Our work demonstrates that the noise floor is
non-trivial: CI widths of 5-7 percentage points on citation share are
common for SearchGPT domains, and improvements of this magnitude or
smaller cannot be reliably attributed to an intervention without
repeated sampling and statistical validation.

\hypertarget{user-behavior-and-attention-in-ai-cited-responses}{%
\subsection{User Behavior and Attention in AI-Cited
Responses}\label{user-behavior-and-attention-in-ai-cited-responses}}

The behavioral consequences of citations in AI-generated responses
represent an adjacent research thread. Joachims et al. \cite{joachims2005} and
Dupret et al. \cite{dupret2008} establish the foundational attention and click
bias models for traditional search, which inform how citation position
may influence user engagement. Cutrell et al. document the role of
snippet quality in click behavior \cite{cutrell2007}. More recently, Ding et al.
examine trust effects of citations in AI responses \cite{ding2025}, and He et
al. study citation UI design and its effects on user attention \cite{he2025}.
Venkit et al. characterize user behavior patterns with answer engines
more broadly \cite{venkit2025}.

Bink et al. investigate credibility perceptions of featured snippets
\cite{bink2022}, and Hu et al. examine inconsistency in AI Overview presence
across repeated queries \cite{hu2025}. The latter is the empirically closest
work to ours: they observe a 33\% inconsistency rate in whether AI
Overviews appear at all for a given query, which is an instance of the
same system-level stochasticity we study. However, Hu et al. treat
inconsistency as a confound to be controlled rather than as the object
of study; we treat it as the primary phenomenon of interest.

\hypertarget{evaluation-variance-and-statistical-rigor}{%
\subsection{\texorpdfstring{ Evaluation Variance and Statistical
Rigor}{ Evaluation Variance and Statistical Rigor}}\label{evaluation-variance-and-statistical-rigor}}

Madaan et al. address variance in Large Language Model (LLM) benchmark
evaluation, arguing that single-number comparisons without variance
reporting are misleading, and introducing variance metrics for benchmark
results \cite{madaan2024}. This work is the closest methodological ally to ours:
both papers argue that point estimates without uncertainty
quantification are insufficient, and both use sampling-based methods to
characterize variability.

The key distinction is scope: Madaan et al. address evaluation variance
in a controlled benchmark setting, where queries and evaluation criteria
are fixed and the only source of variability is the model's stochastic
generation. We address measurement uncertainty in a deployed, production
search system where ground truth is unavailable, the response
distribution is non-stationary, and sampling is the only feasible
approach to characterizing system behavior. The bootstrap methods we
apply follow the tradition of Efron and Tibshirani \cite{efron1993} and are
standard in empirical software engineering and information retrieval
evaluation.

\hypertarget{background-and-definitions}{%
\section{Background and Definitions}\label{background-and-definitions}}

\hypertarget{generative-search-and-citation-structure}{%
\subsection{Generative Search and Citation
Structure}\label{generative-search-and-citation-structure}}

An answer engine is a system that accepts a natural-language query and
returns a synthesized natural-language response, where claims within the
response are attributed to a set of cited web sources. We use the terms
answer engine and generative search engine interchangeably throughout.
Each platform studied here---Perplexity Search, OpenAI SearchGPT, and
Google Gemini---implements this paradigm, though with meaningful
differences in citation volume, citation presentation, and response
format.

A response is a single synthesized answer to a single query. Each
response contains one or more citations, where a citation is a reference
to a specific URL that the platform identifies as a source supporting
the response content. The domain of a citation is the registered domain
of the cited URL (e.g., \emph{nationalgeographic.com} for any URL under
that domain). Citation extraction is platform-specific: Perplexity and
SearchGPT expose citations as structured fields in their API responses
(Appendix A provides illustrative examples), while Gemini embeds source
references in a format requiring text-level parsing. All three platforms
were handled by platform-specific extraction procedures; see Section 4.4
for details.

\hypertarget{visibility-metrics}{%
\subsection{Visibility Metrics}\label{visibility-metrics}}

We define three visibility metrics. Let \emph{S} be a sample of \emph{N}
responses to a set of queries submitted to a single platform. Let
\emph{c}(\emph{d, S}) denote the total number of times domain \emph{d}
is cited across all responses in \emph{S}, and let
\begin{equation}
C(S) = \sum_{d} c(d, S)
\end{equation}
be the total number of citations in \emph{S}.

\textbf{Citation count} for domain \emph{d} in sample \emph{S} is
\emph{c}(\emph{d, S}). Citation count is an unbounded sum that scales
with \emph{N} and with the per-response citation volume of the platform.
It is not directly comparable across platforms or across samples of
different sizes.

\textbf{Citation share} for domain \emph{d} in sample \emph{S} is
\begin{equation}
s(d, S) = \frac{c(d, S)}{C(S)}.
\end{equation}
Citation share normalizes by total citations, making it comparable
across samples and across platforms. The near-perfect correlation
between citation count and citation share within a platform (Spearman
\emph{$\rho$} = 0.994-0.997; Section 5.1) confirms that share preserves the
ordinal information of count while enabling cross-platform comparison.

\textbf{Citation prevalence} for domain \emph{d} in sample \emph{S} is
\begin{equation}
p(d, S) = \frac{\bigl|\{\, r \in S : d \text{ cited in } r \,\}\bigr|}{N}.
\end{equation}
Prevalence measures the fraction of responses that include at least one
citation to \emph{d}, regardless of how many times \emph{d} is cited
within a single response, \emph{r}. Prevalence is sensitive to the
breadth of domain coverage across responses rather than the depth of
citation within any single response.

In industry practice, metrics such as citation count, citation share,
and citation prevalence are typically reported as fixed values
describing domain visibility on a platform. This interpretation is
misleading. Each metric is computed from a finite sample of responses
and therefore represents an estimate of an underlying response
distribution rather than a deterministic property of the system.
Formally, these quantities should be interpreted as sample estimators of
population parameters governing the probability that a domain is cited
by the platform. As a result, the observed values depend on the
particular set of responses collected and are subject to sampling
variability.

All three are therefore sample statistics, functions of a finite
collection of responses, and therefore random variables subject to
sampling variation.

Formally, citation share can be written as the estimator
\begin{equation}
\widehat{s}_{S}(d) = \frac{c(d, S)}{C(S)}
\end{equation}
where \({\widehat{s}}_{S}\left( d \right)\) denotes the estimator of
citation share for domain \emph{d} computed from response sample
\emph{S}. Under this interpretation,
\({\widehat{s}}_{S}\left( d \right)\) estimates the expected citation
share of domain \emph{d} under repeated sampling of responses from the
platform.

This interpretation places citation visibility measurement within the
broader framework of statistical estimation for stochastic systems. A
generative search engine can be viewed as producing responses drawn from
an underlying response distribution conditioned on a query. Visibility
metrics computed from a finite set of queries therefore estimate
properties of that distribution rather than fixed attributes of the
platform. Reliable measurement requires repeated sampling and
uncertainty quantification to characterize the variability of these
estimates. The analyses in this paper adopt this perspective, treating
citation visibility metrics as estimators whose statistical properties
can be studied through repeated sampling.

This perspective reframes AI visibility measurement as a statistical
estimation problem for stochastic information systems rather than a
deterministic ranking analysis.

\hypertarget{system-level-stochasticity-vs.-measurement-uncertainty}{%
\subsection{System-Level Stochasticity vs. Measurement
Uncertainty}\label{system-level-stochasticity-vs.-measurement-uncertainty}}

Because visibility metrics are sample estimators computed from a finite
collection of responses, their observed values can vary across repeated
measurements. Two distinct sources of variability contribute to this
variation. The first arises from the non-deterministic behavior of
generative search systems themselves, while the second arises from the
statistical uncertainty inherent in estimating population quantities
from finite samples. We distinguish these sources as system-level
stochasticity and measurement uncertainty, respectively.

Two distinct sources of variability are present in this study, and they
must not be conflated.

System-level stochasticity refers to the property that the answer engine
itself is non-deterministic: the same query submitted twice produces
different responses, because generation is stochastic (temperature
\textgreater{} 0), and because the retrieval components that ground
citations may return different candidate documents on each call. This is
a property of the system, not of the measurement process.

Measurement uncertainty refers to the property that any finite sample of
responses produces estimates with sampling error. Even if the underlying
response distribution were perfectly stable, a sample of 200 queries
would produce citation share estimates that deviate from the true
population values. This is a property of the measurement process.

Both are present in this study. The high-frequency sampling regime,
queries submitted at ten-minute intervals, helps isolate system-level
stochasticity by minimizing the opportunity for content changes in cited
pages, for example, between samples. Bootstrap confidence intervals
address measurement uncertainty by characterizing the sampling
distribution of the estimates. The content-change validation in Section
6 provides empirical evidence that the variability observed is
predominantly system-level: the underlying pages are largely stable, and
the citation variability reflects engine behavior rather than changing
source material.

A third property, within-sample non-stationarity, is distinct from both.
A citation distribution is stationary if its statistical properties
(mean, variance, and correlation structure) remain constant across the
sequence of queries within a single collection run. Non-stationarity
means the distribution shifts across the query sequence: early queries
in the batch elicit a systematically different citation pattern than
later queries, so the 200 responses in a single sample are not
exchangeable draws from a fixed distribution. The practical consequence
is that confidence intervals built on the stationarity assumption,
including the bootstrap, may understate true uncertainty, and
convergence of CI width with sample size may be non-monotone.
Within-sample non-stationarity is most likely driven by query-ordering
effects: if queries of similar type cluster together in the submission
sequence, the running citation distribution shifts as the composition of
the query batch changes. This is a measurement design problem rather
than a system property, and it is distinct from between-sample
stationarity, whether the citation distribution is the same on different
occasions, which is what the daily time series in Section 5.2 and the
rank stability analysis in Section 5.8 address. Section 5.7 provides
empirical evidence of within-sample non-stationarity for SearchGPT, in
the form of non-monotone CI convergence curves.

The experimental design in Section 4 is structured to isolate these two
sources of variability by combining repeated sampling with
short-interval collections that minimize the probability of content
changes in cited sources.

\hypertarget{experiment-design}{%
\section{Experiment Design}\label{experiment-design}}

\hypertarget{topics}{%
\subsection{Topics}\label{topics}}

Three consumer product topics were selected to represent varied market
structures and information environments. Bird feeders is a fragmented
market with no dominant brand, driven by ecological diversity and
hobbyist communities; the information environment includes niche
specialty sites alongside general natural history publishers. Running
gear is a brand-concentrated, multi-product category spanning footwear,
electronics, and apparel; the information environment mixes commercial
retailers, specialist review sites, and general sports media.
Multivitamins for adults is a health and life sciences context with
strong representation from institutional and clinical sources alongside
commercial and wellness publishers. All three topics are unbranded
queries of broad consumer interest.

\hypertarget{query-generation}{%
\subsection{Query Generation}\label{query-generation}}

For each topic, queries were generated by prompting an LLM (ChatGPT) to
produce queries conditioned on a randomly selected query type drawn from
a fixed set of ten categories: strengths and weaknesses, popular
options, value for money, environmental impact, expert recommendations,
feature comparisons, customer satisfaction, what to look for, common
complaints, and emerging trends. This process was repeated until 200
queries had been generated for each topic. The ten-category structure is
confirmed empirically: facet mapping across all collected responses
yields exactly ten distinct query type values.

Uniqueness was not enforced. The intent was to simulate the distribution
of queries that an answer engine receives from a diverse user
population, in which popular queries recur. If the generating model
produced a repeated query, it was retained on the grounds that
frequently generated queries are likely to correspond to frequently
asked user queries. Table 1 summarizes the resulting query sets,
including total queries, unique queries, and repetition rate per topic.

\begin{table}[htbp]
\centering
\small
\caption{Query generation summary: total queries, unique queries, and repetition rate per topic.}
\label{tab:querygen}
\begin{tabular}{lccc}
\toprule
Topic & Total Queries & Unique Queries & Repetition Rate \\
\midrule
bird feeders & 200 & 172 & 14.0\% \\
multivitamins for adults & 200 & 152 & 24.0\% \\
running gear & 200 & 185 & 7.5\% \\
\bottomrule
\end{tabular}
\end{table}

We acknowledge that query generation by LLM introduces a potential
confound: the generating model may have implicit associations between
topics and sources that shape the query distribution. This is discussed
further in Section 8.

\hypertarget{data-collection}{%
\subsection{Data Collection}\label{data-collection}}

The same 200 queries per topic were submitted to Perplexity Search,
OpenAI SearchGPT, and Google Gemini under two sampling regimes.

Daily samples: queries were submitted once per day over nine consecutive
days, yielding nine samples of approximately 200 responses per topic per
platform. Table 2 provides summary statistics for the responses and
citations by platform and topic. (A daily breakdown of these same
statistics is available in Appendix B.)

\begin{table}[htbp]
\centering
\caption{Summary statistics for the daily sampling regime: responses and citations by topic and platform.}
\label{tab:daily}
\resizebox{\textwidth}{!}{%
\begin{tabular}{llcccccc}
\toprule
\multicolumn{3}{c}{} & \multicolumn{5}{c}{Citations} \\
\cmidrule(lr){4-8}
Platform & Topic & Mean Responses & Mean & Median & Std & Min & Max \\
\midrule
Gemini & bird feeders & 198.8 & 42.6 & 37.0 & 22.7 & 6 & 158 \\
Gemini & multivitamins & 198.6 & 39.6 & 36.0 & 19.6 & 3 & 197 \\
Gemini & running gear & 199.0 & 43.1 & 40.0 & 21.0 & 1 & 202 \\
\addlinespace
SearchGPT & bird feeders & 196.2 & 7.1 & 7.0 & 2.7 & 1 & 18 \\
SearchGPT & multivitamins & 193.9 & 5.9 & 5.0 & 2.0 & 1 & 15 \\
SearchGPT & running gear & 188.1 & 6.6 & 5.0 & 2.6 & 1 & 19 \\
\addlinespace
Perplexity & bird feeders & 199.8 & 22.0 & 20.0 & 9.7 & 2 & 72 \\
Perplexity & multivitamins & 199.9 & 20.0 & 19.0 & 8.3 & 3 & 52 \\
Perplexity & running gear & 199.8 & 22.5 & 22.0 & 8.7 & 4 & 56 \\
\bottomrule
\end{tabular}}
\end{table}

The substantial variation in median citation volume, from 5-7 citations
per response on SearchGPT to 36-40 on Gemini, is immediately visible and
has direct implications for how metrics are defined and compared across
platforms.

High-frequency samples: for the running gear topic, queries were
submitted to Gemini, SearchGPT, and Perplexity at ten-minute intervals
over approximately four hours, yielding 25 samples per platform. The
goal was to minimize the probability of content changes in cited pages
between samples, thereby isolating system-level stochasticity. At
ten-minute intervals, substantive editorial changes to web content are
unlikely for the majority of sources. Table 3 provides summary
statistics for the responses and citations by platform.

\begin{table}[htbp]
\centering
\small
\caption{Summary statistics for the high-frequency sampling regime: responses and citations by platform.}
\label{tab:hf}
\begin{tabular}{llcccccc}
\toprule
\multicolumn{3}{c}{} & \multicolumn{5}{c}{Citations} \\
\cmidrule(lr){4-8}
Platform & Topic & Mean Responses & Mean & Median & Std & Min & Max \\
\midrule
Gemini & running gear & 198.8 & 47.1 & 42.0 & 23.9 & 1 & 213 \\
SearchGPT & running gear & 190.8 & 6.7 & 5.0 & 2.7 & 1 & 40 \\
Perplexity & running gear & 199.9 & 22.7 & 22.0 & 8.8 & 3 & 63 \\
\bottomrule
\end{tabular}
\end{table}

\hypertarget{citation-extraction}{%
\subsection{Citation Extraction}\label{citation-extraction}}

Citations were extracted from raw generative responses using
platform-specific parsing procedures, reflecting differences in how each
platform structures its outputs. Extracted fields include the reference
URL and domain for each cited source, along with the response and query
identifiers necessary for linking citations back to their generating
responses.

After extraction and deduplication, the daily dataset contained 374,052
citations and the high-frequency dataset contained 379,276 citations.
The extraction methodology was consistent across all samples for a given
platform.

\hypertarget{content-change-monitoring}{%
\subsection{\texorpdfstring{ Content-Change
Monitoring}{ Content-Change Monitoring}}\label{content-change-monitoring}}

As a methodological control, the HTML of cited URLs was scraped and
stored during each data collection job. Human-readable content was
extracted from raw HTML using Trafilatura, and a SHA-256 checksum was
computed over the extracted text. Checksums were stored alongside
scraping metadata in a database indexed by URL and job identifier,
enabling detection of content changes between samples without requiring
re-retrieval and re-comparison of full HTML. Scraping coverage is
subject to server-side blocking (e.g., Cloudflare, CAPTCHA challenges)
and to format limitations (video content and PDF documents, for example,
cannot be processed by Trafilatura). Results of the content-change
validation are reported in Section 6.

\hypertarget{findings}{%
\section{Findings}\label{findings}}

The findings are organized from individual responses to aggregate
visibility metrics. We first show that variability is visible at the
response level (Section 5.1), then characterize aggregate visibility
over time (Section 5.2). Next, we differentiate between frequently and
infrequently cited domains and establish how they will be handled in
this analysis (Section 5.3). We establish the distributional foundation,
power-law structure (Section 5.4) and log-space dispersion (Section
5.5), before demonstrating the application of bootstrap confidence
intervals (Section 5.6) and analyzing the effect of sample size on
interval width (Section 5.7). Finally, we close with a distribution-wide
assessment of rank stability across samples (Section 5.8).

\hypertarget{variability-in-individual-responses}{%
\subsection{Variability in Individual
Responses}\label{variability-in-individual-responses}}

Before aggregating across queries, we examine whether variability is
visible at the individual response level. For the same query submitted
across multiple samples, the answer engines returned different sets of
cited sources. Table 4 summarizes the response-level citation overlap
structure: rows represent the mode number of citations per response for
each platform, and columns represent the count of matching citations
across repeated runs of the same query. Matching is measured at the
domain level (not URL level): two responses match on a domain if both
cite at least one URL from that domain. URL-level matching would be
strictly more variable

\begin{table}[htbp]
\centering
\caption{Response-level citation matching. Rows give per-platform Jaccard overlap statistics across repeated runs of the same query, including identical citation rate, zero-overlap rate, and median Jaccard index. Matching is at the domain level.}
\label{tab:jaccard}
\resizebox{\textwidth}{!}{%
\begin{tabular}{llccccccc}
\toprule
Platform & Topic & Queries & Pairs & Median Jaccard & Identical Rate & Zero Overlap Rate & Mean Intersection & Mean Unique Domains \\
\midrule
Gemini & bird feeders & 189 & 6{,}804 & 0.31 & 0.07 & 0.35 & 4.90 & 10.54 \\
Gemini & multivitamins & 192 & 6{,}912 & 0.29 & 0.01 & 0.80 & 4.49 & 10.55 \\
Gemini & running gear & 192 & 6{,}912 & 0.29 & 0.10 & 1.62 & 4.73 & 10.88 \\
\addlinespace
SearchGPT & bird feeders & 180 & 6{,}480 & 0.40 & 7.25 & 6.08 & 2.50 & 4.38 \\
SearchGPT & multivitamins & 180 & 6{,}480 & 0.38 & 8.02 & 7.19 & 2.03 & 3.76 \\
SearchGPT & running gear & 169 & 6{,}084 & 0.33 & 5.34 & 8.93 & 2.04 & 4.04 \\
\addlinespace
Perplexity & bird feeders & 198 & 7{,}128 & 0.50 & 3.93 & 2.40 & 3.32 & 5.24 \\
Perplexity & multivitamins & 199 & 7{,}164 & 0.50 & 4.65 & 1.05 & 3.34 & 5.11 \\
Perplexity & running gear & 198 & 7{,}128 & 0.50 & 3.25 & 1.07 & 3.49 & 5.31 \\
\bottomrule
\end{tabular}}
\end{table}

Platform-level Jaccard medians are 0.29-0.31 for Gemini, 0.33-0.40 for
SearchGPT, and 0.50 for Perplexity across all three topics. The
identical citation rate, where two responses to the same query share
every cited source, ranges from near-zero for Gemini (0.01-0.10\%) to
3-8\% for SearchGPT and Perplexity. The zero-overlap rate is highest for
SearchGPT (6-9\%) and lowest for Perplexity (1-2\%), consistent with the
bimodal behavior shown below.

\begin{figure}[htbp]
\centering
\includegraphics[width=\textwidth,height=0.9\textheight,keepaspectratio]{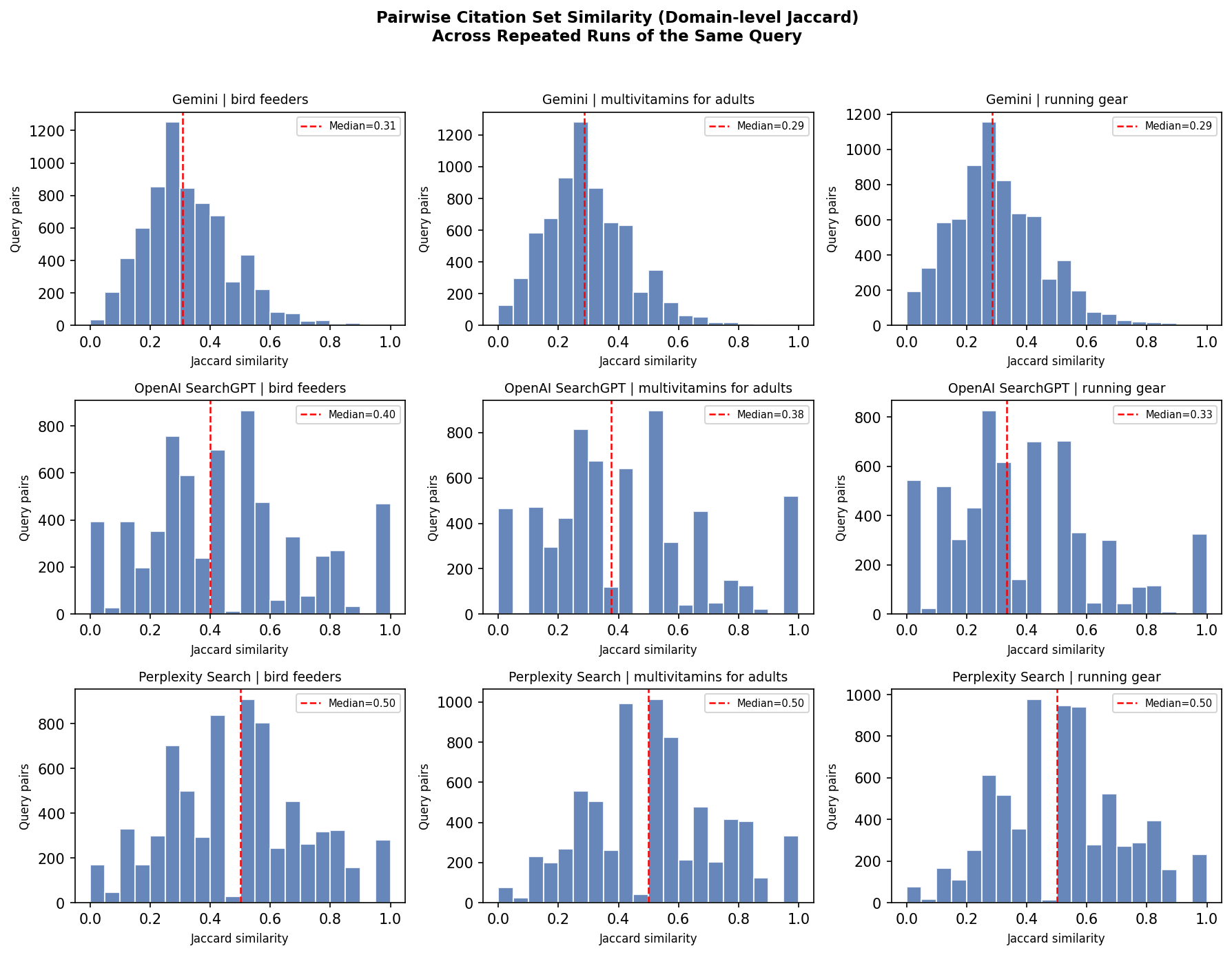}
\caption{Pairwise citation set similarity (domain-level Jaccard) across repeated runs of the same query: rows = platform, columns = topic.}
\label{fig:jaccard}
\end{figure}

The pairwise Jaccard distribution histograms reveal the structure behind
these summaries. Gemini's distributions are broad and unimodal, peaking
near 0.30-0.35 with a slight skew to the right. The platform rarely
produces either complete overlap (Jaccard = 1) or complete divergence
(Jaccard = 0).

The SearchGPT distributions are substantially broader and more irregular
than Gemini's. Rather than forming a single unimodal peak, the
histograms span nearly the full 0-1 range and exhibit multiple local
peaks. This pattern indicates that repeated runs of the same query
frequently produce distinct citation sets, with similarity varying from
minimal overlap to near identity. The multimodal structure suggests
regime-like behavior in the citation distribution rather than sampling
from a single stable pool of sources.

The Perplexity distributions occupy an intermediate position between
Gemini and SearchGPT. They are centered at substantially higher
similarity levels (median $\approx$ 0.50 across topics), indicating that
repeated runs tend to share roughly half of their cited domains. The
distributions remain moderately broad and show mild multimodality,
suggesting that Perplexity responses draw from a stable core of
frequently cited domains while alternating among a small number of
secondary citation sets. This behavior contrasts with Gemini's smoother
unimodal distributions centered near 0.30 and with SearchGPT's highly
irregular, multi-peaked distributions that span nearly the full 0-1
range. A second analysis plots median Jaccard against the modal citation
count per response within each platform.

\begin{figure}[htbp]
\centering
\includegraphics[width=\textwidth,height=0.9\textheight,keepaspectratio]{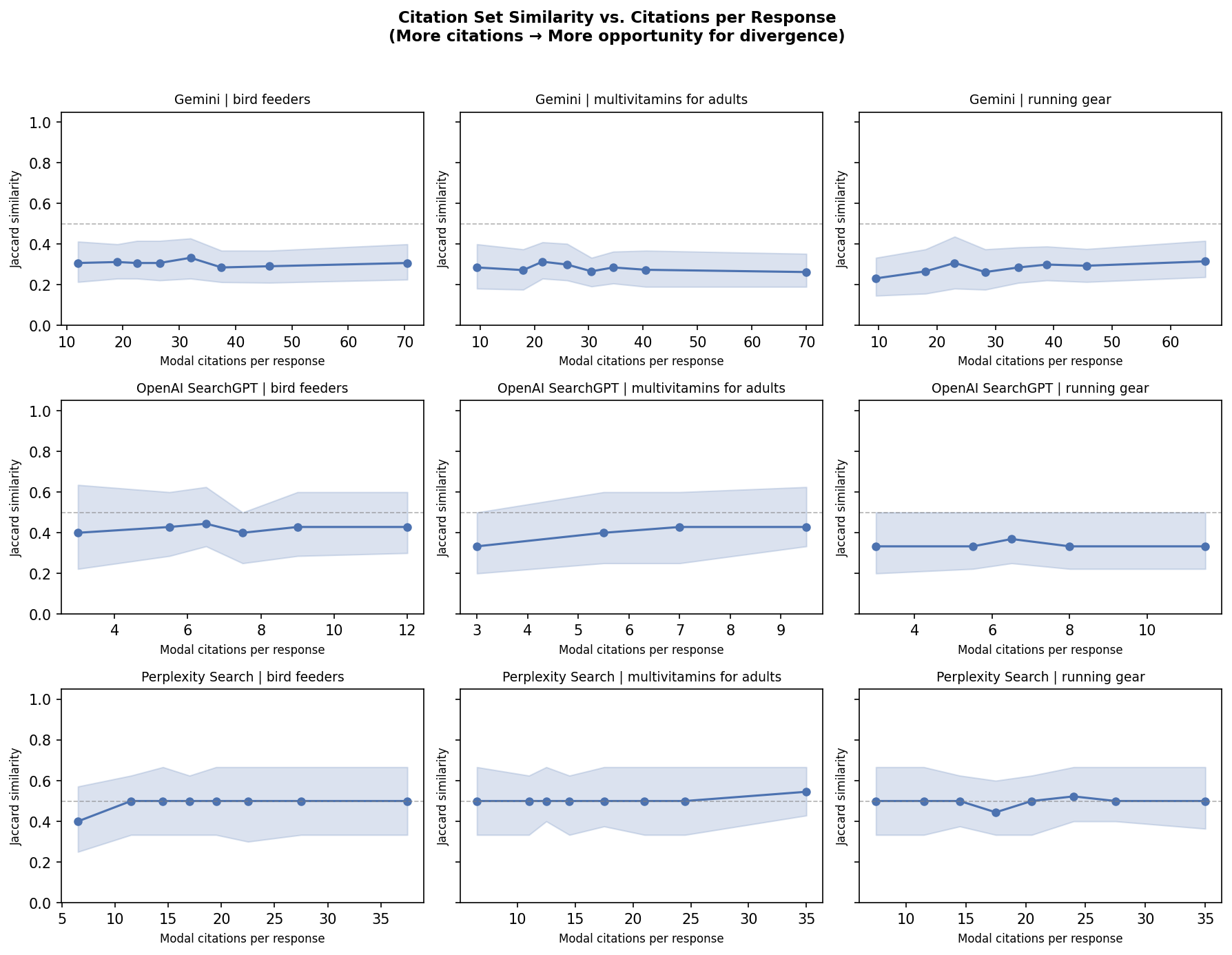}
\caption{Citation set similarity vs.\ citations per response: rows = platform, columns = topic. Jaccard similarity is largely independent of the number of citations.}
\label{fig:jaccard_count}
\end{figure}

The number of citations in a response might plausibly affect
citation-set similarity. In principle, responses that contain more
citations provide more opportunities for divergence across runs: each
additional citation could be replaced by a different source, potentially
reducing overlap between citation sets. If this mechanism dominated,
Jaccard similarity would decline as the number of citations per response
increases.

Figure~\ref{fig:jaccard_count} shows that this effect is largely absent. Across all three
platforms and topics, Jaccard similarity remains essentially flat across
the observed range of citation counts. Gemini maintains similarity near
0.30 across responses containing roughly 10 to 70 citations. SearchGPT
remains near 0.40-0.42 across 3 to 12 citations, while Perplexity holds
near 0.50 across 5 to 35 citations. The fitted lines show little
systematic slope, and the uncertainty bands overlap broadly across
citation-count bins.

The stability of these curves indicates that citation-set similarity is
largely independent of the number of citations in a response. Responses
with many citations are not materially more divergent across runs than
responses with few citations. Instead, the dominant determinant of
citation consistency is platform identity. Each system exhibits a
characteristic level of overlap across repeated runs (approximately 0.30
for Gemini, 0.40 for SearchGPT, and 0.50 for Perplexity), regardless of
how many sources appear in individual responses.

This result suggests that citation variability arises primarily from
platform-level retrieval and ranking behavior, rather than from simple
combinatorial effects associated with response length. In other words,
larger citation sets do not inherently produce more unstable citation
patterns; the consistency of those sets is determined mainly by how each
system constructs and selects its evidence base.

Importantly, this pattern holds despite substantial variation in
citation counts within each platform. If citation-set similarity were
primarily driven by the combinatorics of set size, we would expect
similarity to increase with the number of citations per response. The
absence of such a trend indicates that the observed differences in
similarity are not a mechanical artifact of response length but instead
reflect platform-specific citation selection behavior.

\hypertarget{aggregate-visibility-metrics}{%
\subsection{Aggregate Visibility
Metrics}\label{aggregate-visibility-metrics}}

\begin{figure}[htbp]
\centering
\includegraphics[width=\textwidth,height=0.9\textheight,keepaspectratio]{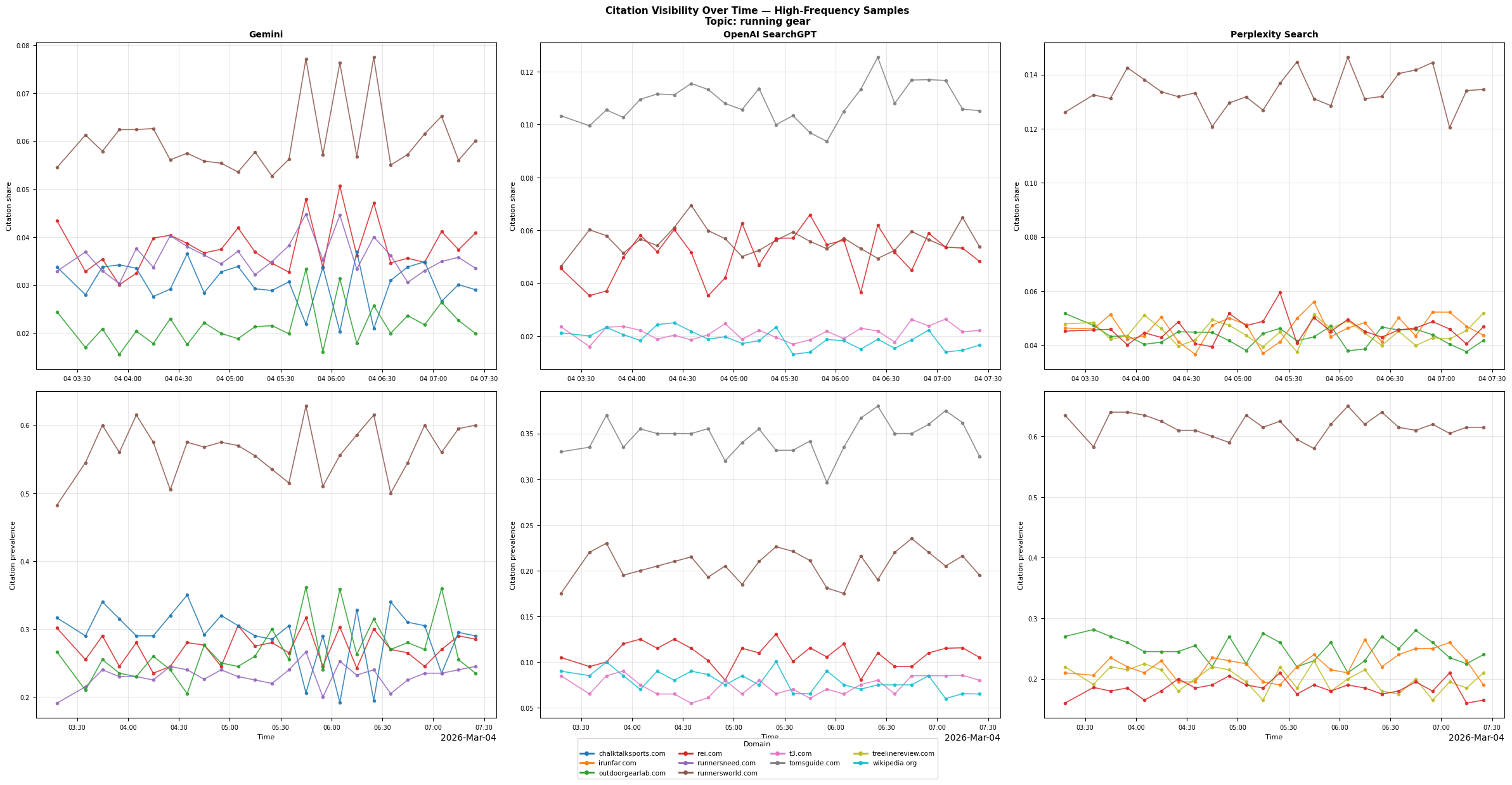}
\caption{Citation visibility over time: high-frequency samples. Topic: running gear. Top row: citation share; bottom row: citation prevalence. Left: Gemini (25 samples, 2026-Mar-04); center: SearchGPT (25 samples, same window); right: Perplexity (25 samples, same window). Top-ranked domains shown as colored lines.}
\label{fig:hf_timeseries}
\end{figure}

We now move from individual responses to aggregated visibility metrics
across all queries in a sample. Figure~\ref{fig:hf_timeseries} shows the high-frequency time
series for the running gear topic, plotting citation share (top row) and
citation prevalence (bottom row) for the top-ranked domains on each
platform across consecutive ten-minute samples.

Because this window is far too short for meaningful changes in the
underlying web content, variation in these series reflects instability
in the platforms' citation selection rather than changes in the source
material itself.

The three platforms exhibit distinct stability regimes. On Gemini,
\emph{runnersworld.com} is the most cited domain, with citation share
fluctuating between 5.3\% and 7.7\%. The remaining top domains exhibit
substantial short-interval volatility, particularly in the latter half
of the series where citation shares oscillate sharply. The prevalence
series mirrors this pattern: several domains show abrupt rises and drops
over adjacent samples, indicating frequent reshuffling of the citation
set rather than smooth sampling from a stable distribution.

On SearchGPT, \emph{tomsguide.com} is the top-cited domain with a
citation share of roughly 9.4-12.5\%, which remains moderately stable
across the sample window. Instability appears primarily among the
lower-ranked domains, which tend to swap positions in pairs---most
visibly between the second and third and fourth and fifth positions.
This pattern suggests that SearchGPT maintains a relatively stable
leading source while the remainder of the citation set alternates among
a small group of competing domains.

Perplexity shows the most stable behavior overall.
\emph{runnersworld.com} again appears as the dominant domain with
citation share between 12.0\% and 14.6\%, and both the share and
prevalence series remain comparatively smooth across the window. The
lower-ranked domains cluster at similar citation levels and occasionally
exchange rank order, but the amplitude of these fluctuations is small
relative to the other platforms. This pattern suggests a stable core
citation set with minor rotation among secondary sources.

Taken together, the platforms form a clear ordering in short-interval
stability: Perplexity is the most stable, followed by SearchGPT, with
Gemini exhibiting the greatest volatility. This ordering mirrors the
patterns observed in the daily sampling regime and in the log-space
dispersion analysis presented in subsequent sections. The consistency of
this ranking across time scales indicates that the stability differences
are structural properties of each engine's retrieval and
citation-selection process, rather than artifacts of the sampling
procedure.

The prevalence series closely track the behavior seen in citation share,
confirming that the volatility reflects changes in which domains appear
in responses rather than small fluctuations in citation counts within
otherwise stable citation sets.

\begin{figure}[htbp]
\centering
\includegraphics[width=\textwidth,height=0.9\textheight,keepaspectratio]{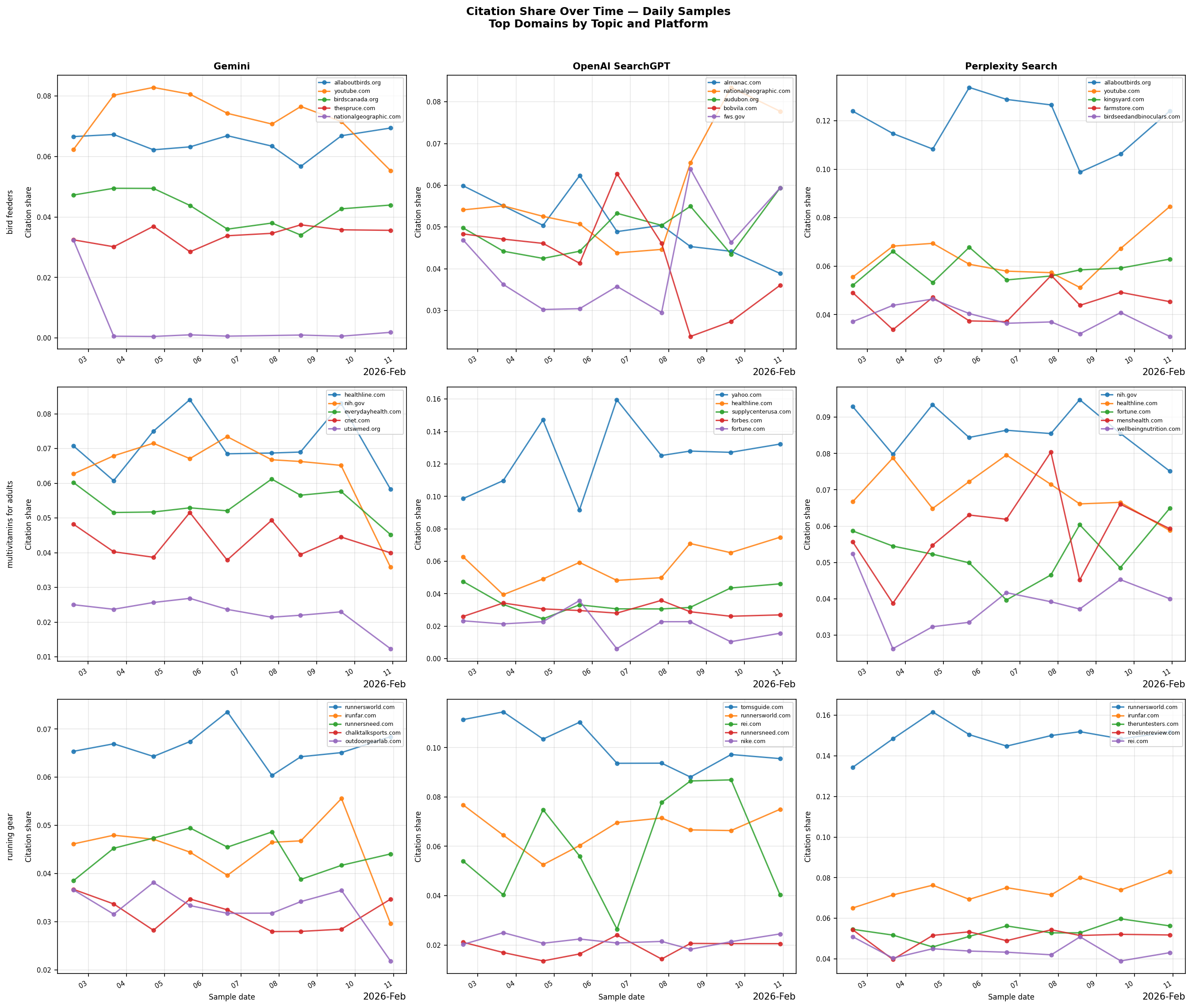}
\caption{Citation share over time: daily samples. Top domains by topic and platform. $3\times3$ grid: rows = topics (bird feeders, multivitamins for adults, running gear), columns = platforms (Gemini, SearchGPT, Perplexity). Top 5 domains per panel shown as colored lines. Samples span Feb 3--11, 2026.}
\label{fig:daily_timeseries}
\end{figure}

Figure~\ref{fig:daily_timeseries} presents the daily time series across all nine topic-platform
panels, confirming that the patterns seen in the high-frequency window
persist at daily resolution and over the nine-day window.

A key observation is that domain rankings by citation share are unstable
across samples. Here, the most anomalous behavior is
\emph{nationalgeographic.com} in the Gemini bird feeders panel: the
domain registers a citation share of approximately 0.032 on day 1 and
reverts to near-zero for all subsequent days. In the SearchGPT
multivitamins panel, yahoo.com citation share ranges from approximately
0.092 to 0.160 within the nine-day window---a factor of nearly 2$\times$ for
the top-ranked domain. The Perplexity bird feeders panel shows
\emph{allaboutbirds.org} swinging between 0.099 and 0.134 across the
nine days but starting and ending at 0.124.

\begin{figure}[htbp]
\centering
\includegraphics[width=\textwidth,height=0.9\textheight,keepaspectratio]{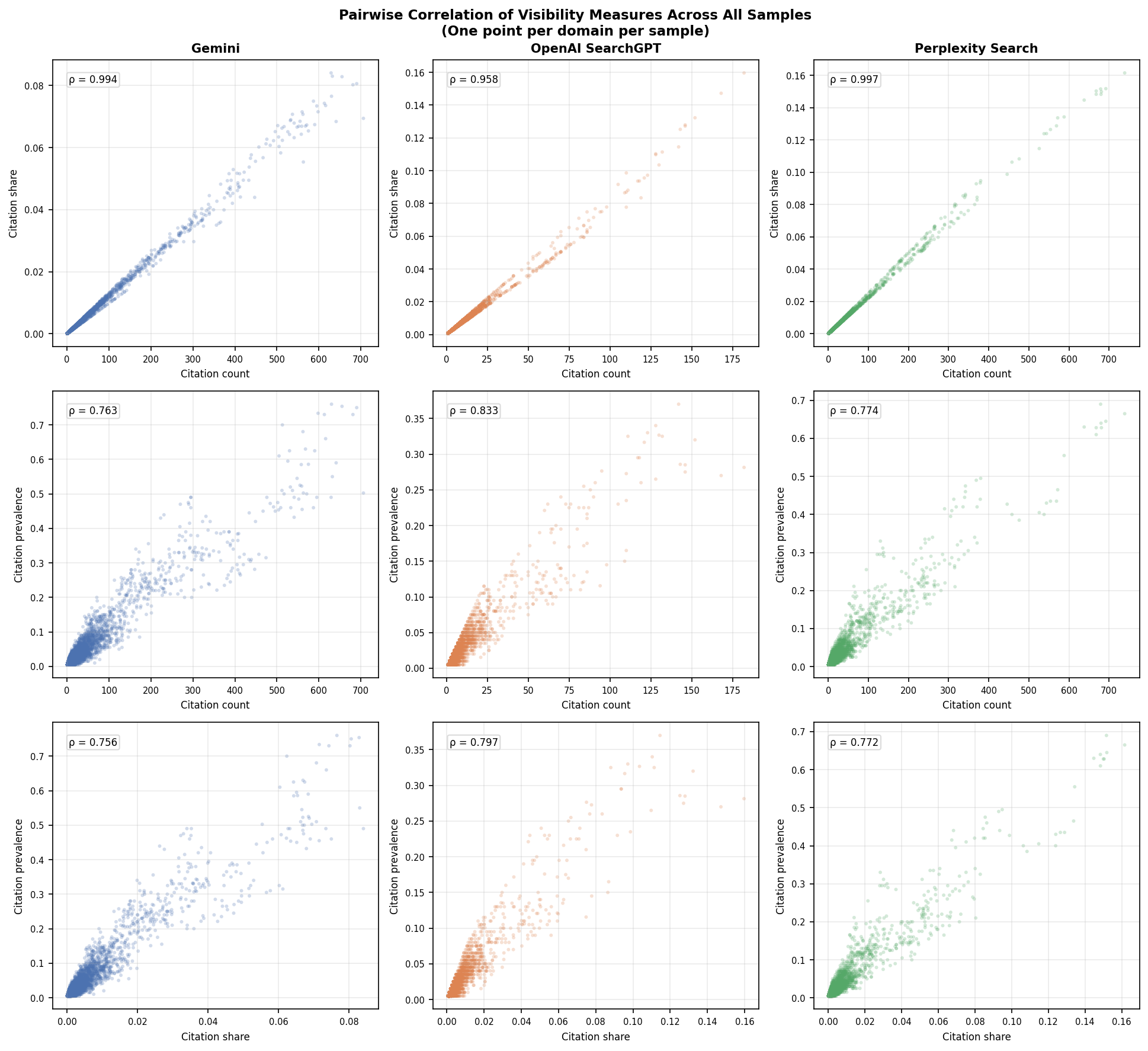}
\caption{Correlation structure of visibility metrics. $3\times3$ scatter grid: rows = metric pairs (citation count vs.\ share; citation count vs.\ prevalence; citation share vs.\ prevalence), columns = platforms. One point per domain per sample. Spearman $\rho$ annotated per panel.}
\label{fig:correlation}
\end{figure}

Figure~\ref{fig:correlation} quantifies the relationships among the three visibility metrics
across all samples.

Citation count and citation share are nearly perfectly correlated within
platform (\emph{$\rho$} = 0.958-0.997), confirming that share preserves the
ordinal information of count while enabling cross-platform comparison.
The x-axis scales differ substantially across platforms
(Gemini/Perplexity: 0-700 citations; SearchGPT: 0-175), making the
volume difference immediately visible and confirming that raw counts
cannot be compared cross-platform. Citation prevalence is moderately
correlated with both count and share (\emph{$\rho$} = 0.756-0.833), with
substantial scatter at higher values. The fan-shaped scatter in the
share-prevalence panels confirms that prevalence is structurally
distinct from share and captures a different dimension of visibility: a
domain can have high share by appearing many times in a few responses,
or high prevalence by appearing once in many responses.

As shown in Table 2, median citations per response differ substantially
across platforms: 36-40 for Gemini, 19-22 for Perplexity, and 5-7 for
SearchGPT. These volume differences mean that raw citation counts are
not comparable across platforms; citation share is the appropriate
primary metric for cross-platform analysis.

\hypertarget{infrequently-cited-domains}{%
\subsection{Infrequently Cited
Domains}\label{infrequently-cited-domains}}

Domains that are not cited in every sample present a distinct
measurement challenge. When a domain is absent from a sample, its
measured visibility is zero, but this does not imply that its citation
probability is zero. The domain may simply not have been sampled in that
run. Standard bootstrap methods designed for continuously distributed
data do not apply directly to these zero-inflated observations.

\begin{figure}[htbp]
\centering
\includegraphics[width=\textwidth,height=0.9\textheight,keepaspectratio]{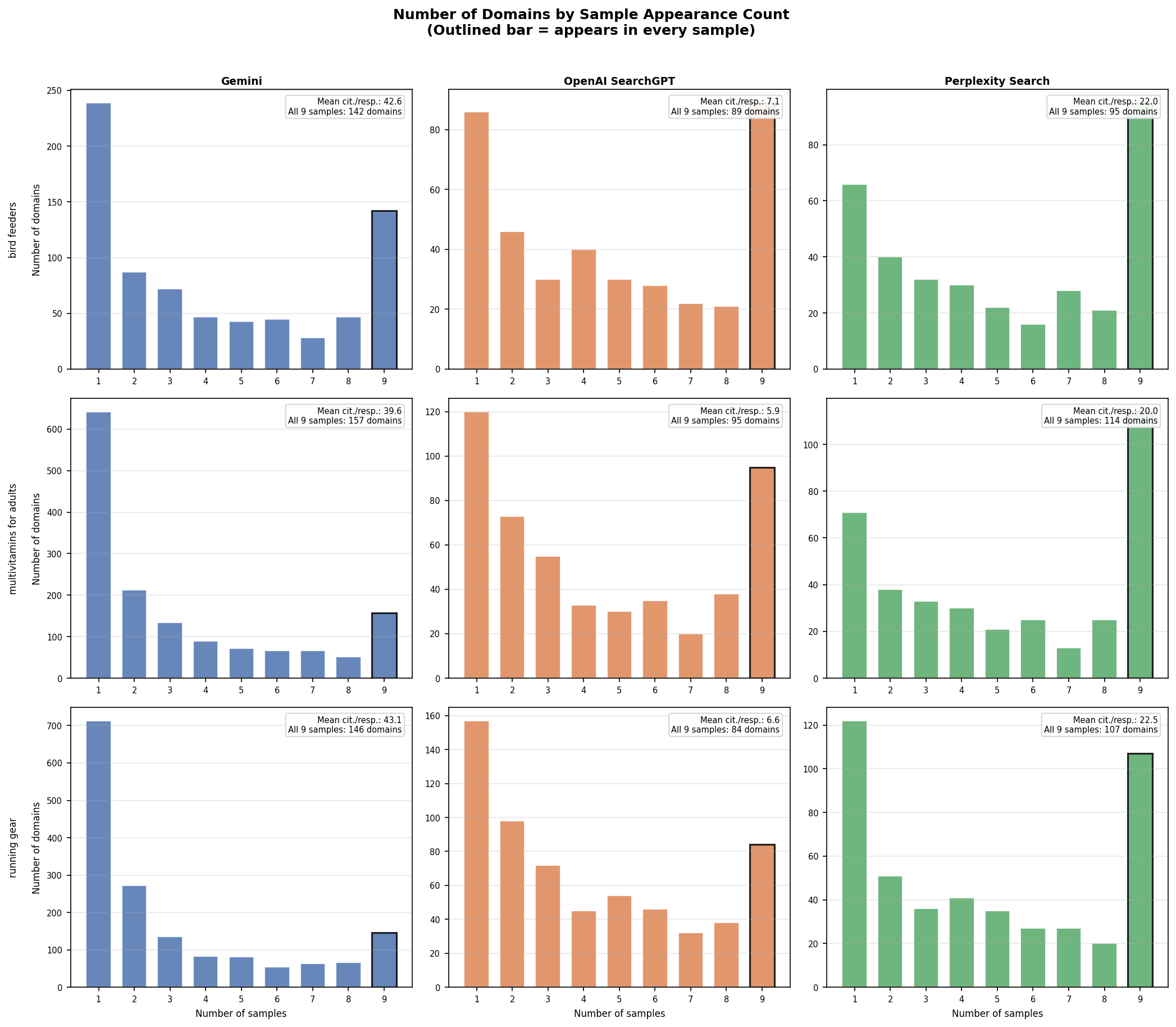}
\caption{Domain count by number of samples cited. $3\times3$ grid: rows = topics, columns = platforms. Bar charts showing the number of domains appearing in exactly 1, 2, \dots, 9 of the nine daily samples. Outlined bar marks the all-9-samples bin (frequently cited domains).}
\label{fig:domain_count}
\end{figure}

For the purposes of this analysis only, a domain is classified as
frequently cited if it appears in every sample for a given platform and
topic. Figure~\ref{fig:domain_count} plots the empirical distribution of domain appearance
counts across the nine daily samples, one panel per platform-topic
combination.

The distributions exhibit a predominantly U-shaped pattern, with
platform variation. For Gemini, the U-shape is pronounced: the
single-appearance bin dominates (239 domains for bird feeders, 642 for
multivitamins, 713 for running gear), middle bins are substantially
lower, and the all-nine-samples bin is large and clearly separated (142,
157, and 146 domains, respectively). Proportionally, the gap between the
8-bin and the 9-bin is especially sharp---roughly three times as many
domains appear in all nine samples as in exactly eight---providing
strong empirical justification for the threshold. For Perplexity, a
clear U-shape is also present, with the all-nine-samples bin the largest
in two of three topics. For SearchGPT, the distribution is more
J-shaped: the single-appearance bin is highest, but the all-nine bin is
nearly as high, with a comparatively flat middle.

If our result set had contained only a single sample, the distinction
between frequently and infrequently cited domains would have been
substantially harder to draw. Both types of domain can appear at any
response ordinal during collection. The difference is that frequently
cited domains, being structurally embedded in the engine's response
distribution, tend to establish themselves early and consistently;
infrequently cited domains may appear once and never recur. But in a
single sample there is no recurrence data to confirm this, and a domain
that appears in the first ten responses of a single run cannot be
distinguished from one that would appear in every run. Determining the
minimum number of samples, or the minimum number of queries within a
single run, required to reliably classify domains as frequently or
infrequently cited is an open question directly connected to the minimum
sample size guidance identified as a priority in Section 10.

\hypertarget{power-law-structure-of-citation-distributions}{%
\subsection{Power-Law Structure of Citation
Distributions}\label{power-law-structure-of-citation-distributions}}

The time series in Section 5.2 illustrate that visibility metrics vary
across samples, but illustrating variance is not the same as quantifying
it. To quantify uncertainty in citation share and prevalence estimates,
we need to understand the underlying distribution from which those
observations are drawn. Two properties of that distribution are
particularly consequential: its shape, which determines how variance
scales with citation share across the domain ranking; and its
dispersion, which captures how spread out individual domain shares are
across repeated samples. Sections 5.4 and 5.5 characterize each in turn,
providing the distributional foundation for the bootstrap confidence
intervals in Section 5.6.

\begin{figure}[htbp]
\centering
\includegraphics[width=\textwidth,height=0.9\textheight,keepaspectratio]{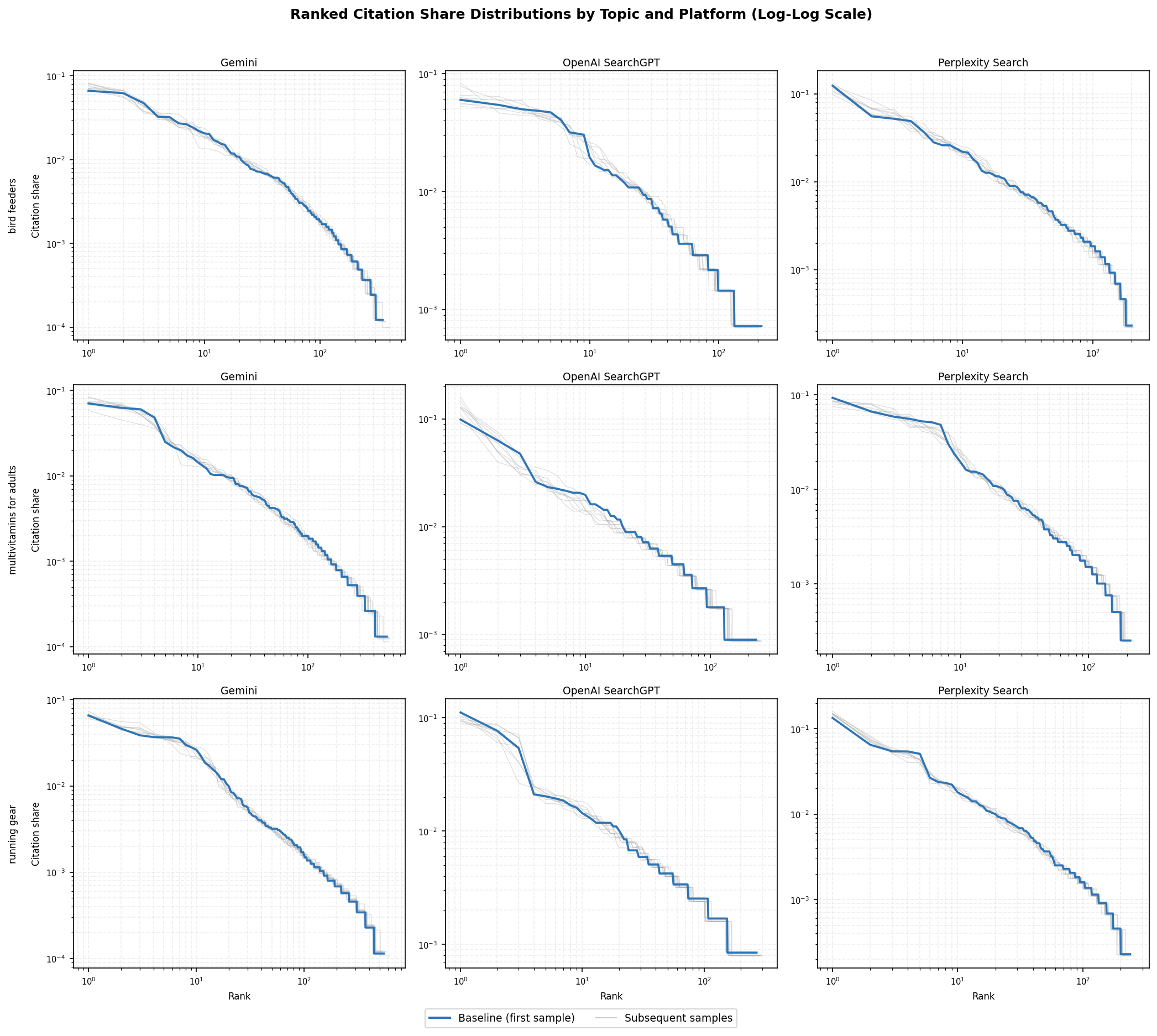}
\caption{Ranked citation share distributions on log-log scale. $3\times3$ grid: rows = topics (bird feeders, multivitamins for adults, running gear), columns = platforms. Thick blue line: baseline (first sample). Thin gray lines: subsequent samples. Log-log scale reveals power-law form.}
\label{fig:ranked}
\end{figure}

Figure~\ref{fig:ranked} plots ranked citation share distributions on a log-log scale
for all nine platform-topic combinations. The baseline (first sample) is
shown as a thick blue line; all subsequent samples as thin gray lines.

All nine panels confirm a clean power-law structure: citation share
decays smoothly across ranks on log-log axes throughout the full domain
range. The distribution shape is highly stable across samples: the gray
lines cluster tightly around the baseline in all panels, with no sample
producing a qualitatively different shape.

The degree of clustering, however, differs by platform in a way that
anticipates the convergence findings in Section 5.7. Gemini and
Perplexity panels show tight inter-sample agreement throughout the full
rank range. SearchGPT panels show noticeably more spread, particularly
in the mid-rank region: the gray lines fan out around the baseline
rather than overlapping it, indicating that individual domain shares
shift more substantially between samples even while the overall
distributional shape is preserved. This inter-sample spread is a
distributional signature of the same within-sample non-stationarity that
produces non-monotonic convergence curves for SearchGPT in Section
5.7---the citation distribution is not merely noisier for SearchGPT but
less stable across the query sequence.

Tail lengths differ by platform: Gemini extends to rank approximately
300-400 (consistent with its higher per-response citation volume),
Perplexity to rank approximately 100-200, and SearchGPT to rank
approximately 100-150. The baseline and subsequent sample curves overlap
closely in every panel, indicating that the distributional shape is
stable even when individual domain ranks shift between samples.

The power-law structure has a direct implication for how variability
should be measured. Because citation share spans several orders of
magnitude across the domain distribution, absolute variance is not
scale-invariant: a domain with 10\% share has far larger absolute
variance than a domain with 0.1\% share, even if their relative
variability is similar. This motivates the use of log-space dispersion
measures in Section 5.5.

\begin{figure}[htbp]
\centering
\includegraphics[width=\textwidth,height=0.9\textheight,keepaspectratio]{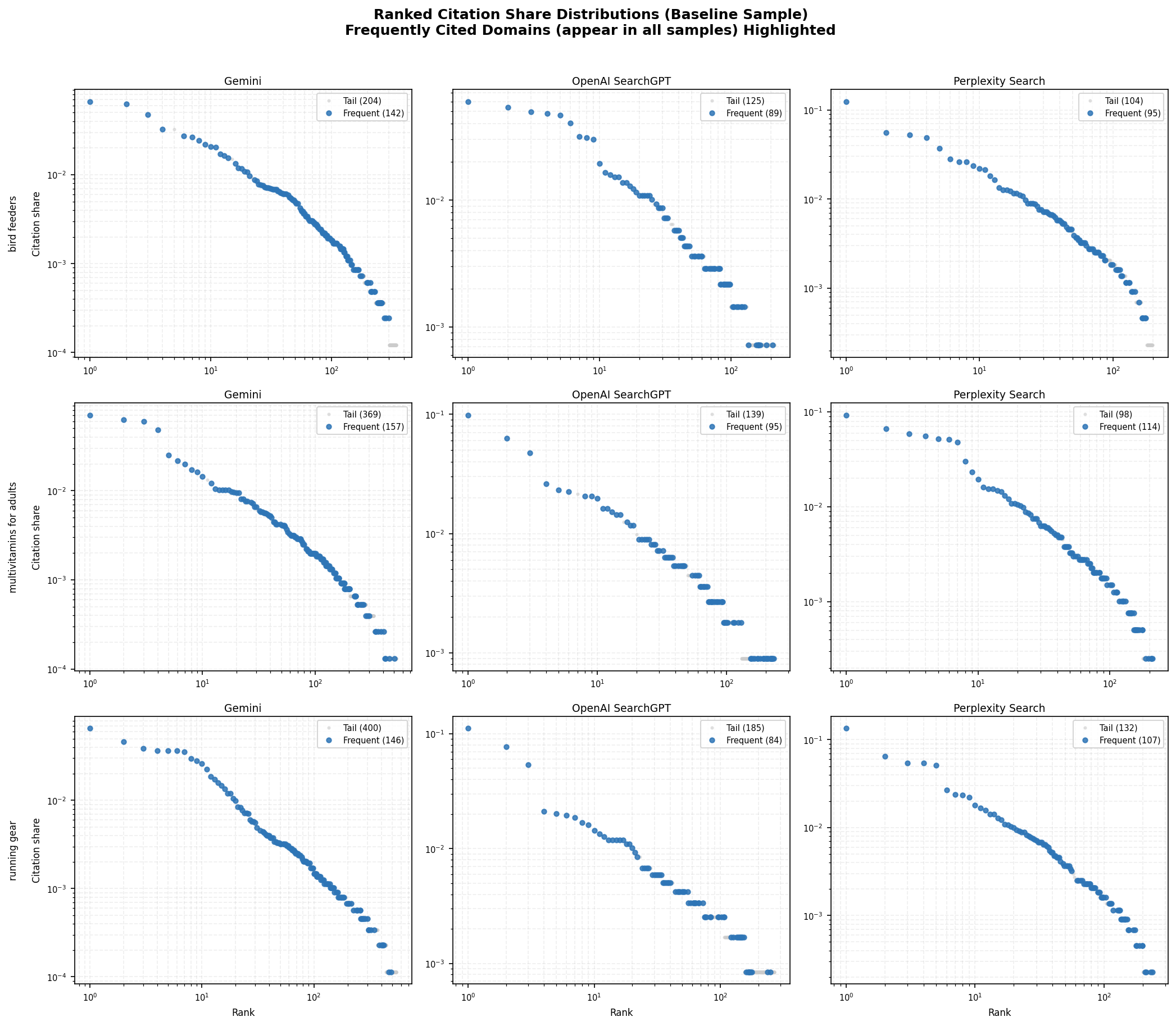}
\caption{Ranked citation share distributions with frequently-cited domain classification. $3\times3$ grid: rows = topics, columns = platforms. Baseline sample only. Filled circles: frequently-cited domains (cited in all 9 samples). Faint points: tail domains. Domain counts per panel annotated in legend.}
\label{fig:ranked_class}
\end{figure}

Figure~\ref{fig:ranked_class} overlays the frequently-cited and infrequently-cited domains on
the log-log rank distribution, confirming that frequently-cited domains
occupy the smooth upper portion of the power-law curve, while tail
domains form the scattered lower portion but follow the same
distributional shape.

\hypertarget{log-space-dispersion}{%
\subsection{Log-Space Dispersion}\label{log-space-dispersion}}

For the set of frequently cited domains, we quantify dispersion using
log-space statistics. Because citation share distributions follow a
power-law form, variability is multiplicative rather than additive:
differences between samples are better characterized as proportional
changes than as absolute differences. The primary dispersion measure
used throughout is the log-std: the standard deviation of log(citation
share) across the nine daily samples, computed per domain. A log-std of
0.5 means that citation share values for a given domain typically
deviate by $\pm$0.5 log units from the geometric mean across samples,
equivalent to a multiplicative factor of approximately $\sqrt{e}$ $\approx$ 1.65,
meaning the share tends to vary between roughly 60\% and 165\% of its
central value. Log-std is scale-invariant: a domain at 1\% share and a
domain at 10\% share with the same log-std are equally variable in
relative terms.

\begin{figure}[htbp]
\centering
\includegraphics[width=\textwidth,height=0.9\textheight,keepaspectratio]{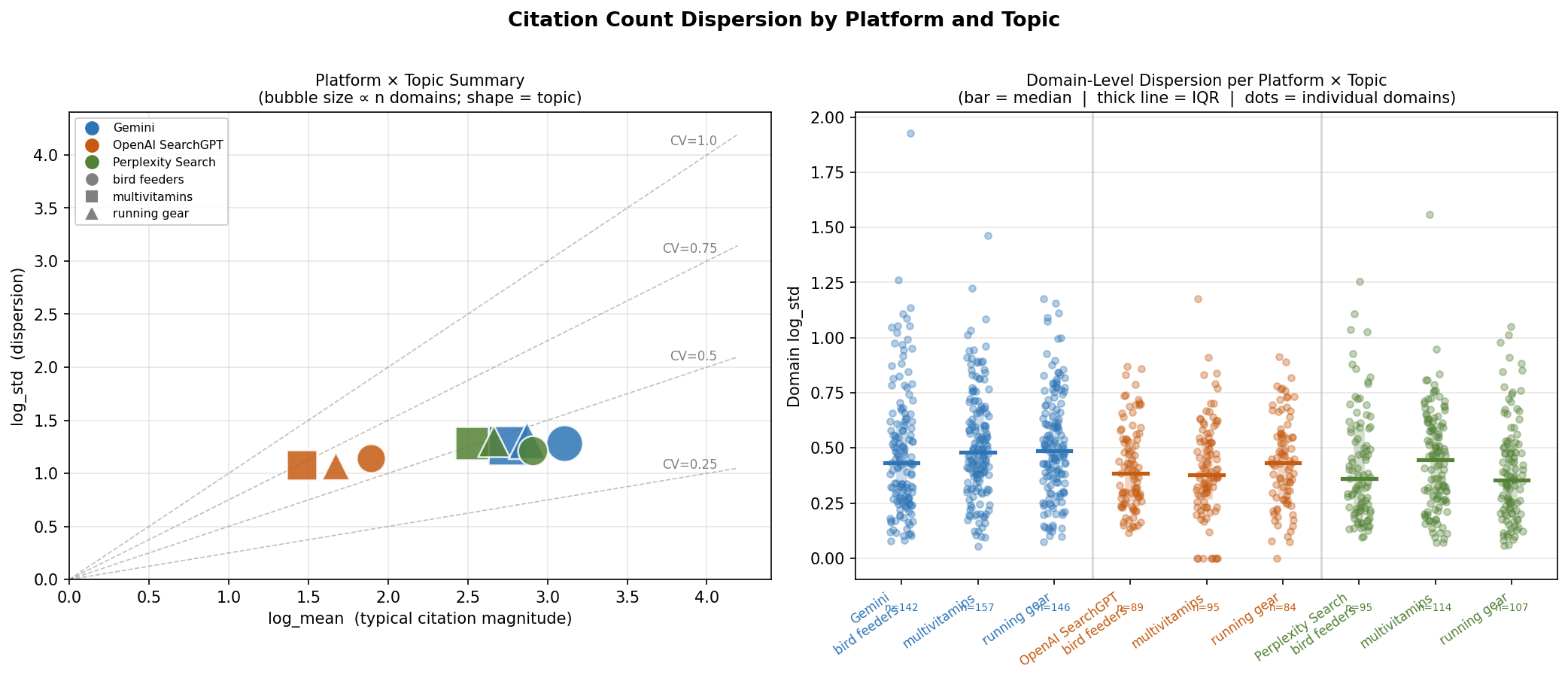}
\caption{Log-space dispersion summary. Panel A: platform-topic summary of mean and median log-std. Panel B: domain-level log-std strip plot, one strip per platform-topic combination (9 strips).}
\label{fig:dispersion}
\end{figure}

Figure~\ref{fig:dispersion} summarizes domain-level log-space dispersion across all nine
platform-topic combinations.

\begin{table}[htbp]
\centering
\small
\caption{Platform-level log-std summary. Mean and median domain log-std aggregated across all topics and frequently-cited domains, by platform.}
\label{tab:logstd}
\begin{tabular}{lcc}
\toprule
Platform & Mean Domain Log-std & Median Domain Log-std \\
\midrule
Gemini & 0.504 & 0.469 \\
SearchGPT & 0.417 & 0.397 \\
Perplexity & 0.421 & 0.378 \\
\bottomrule
\end{tabular}
\end{table}

Platforms differ meaningfully in their dispersion profiles, though the
ordering is counterintuitive. Gemini has the highest mean domain-level
log-std (0.504, median 0.469), followed by Perplexity (mean 0.421,
median 0.378) and SearchGPT (mean 0.417, median 0.397). This does not
mean Gemini is the hardest platform to measure: it reflects that Gemini
has substantially more frequently-cited domains (142-157 per topic) than
SearchGPT (84-95) or Perplexity (95-114), and that the larger tail
includes many low-share domains cited sporadically, which elevates the
platform mean.

For the head of the distribution, the stability ordering reverses. The
most stable frequently-cited domains in the dataset are Perplexity's top
performers: \emph{runnersworld.com} (log-std = 0.062), \emph{nih.gov}
(0.072), \emph{allaboutbirds.org} (0.095). These reflect the structural
concentration that makes dominant domains easy to estimate precisely.
The least stable domains are tail entries that are cited frequently but
erratically: \emph{gardeningknowhow.com} (Gemini bird feeders, log-std =
1.93) and \emph{mindbodygreen.com} (Perplexity multivitamins, log-std =
1.56) are extreme examples---equivalent to citation shares varying by
factors of approximately 6$\times$ and 3$\times$ their geometric mean, respectively.

A notable anomaly appears in SearchGPT multivitamins: nine
frequently-cited domains have log-std = 0.0, meaning identical citation
counts across all nine jobs. This directly corroborates the platform's
Jaccard distribution observed in Section 5.1. SearchGPT has a
deterministic layer that fires consistently for certain domain-query
pairings, almost certainly domains that are associated with a single
query type, which SearchGPT answers identically every time.

The combination of these two properties, a deterministic layer for some
domains and high volatility for others, makes SearchGPT qualitatively
different from Gemini and Perplexity as a measurement target. For Gemini
and Perplexity, higher log-std values reflect quantitative imprecision:
more queries reduce uncertainty at a predictable rate. For SearchGPT,
the volatility in non-deterministic domains reflects within-sample
non-stationarity: the citation distribution shifts across the query
sequence in a way that is not simply smoothed away by collecting more
queries. This distinction has direct consequences for sample size
requirements, which are substantially larger for SearchGPT and in some
cases cannot be met within a realistic budget, a finding elaborated in
Section 5.7.

\hypertarget{bootstrap-confidence-intervals}{%
\subsection{Bootstrap Confidence
Intervals}\label{bootstrap-confidence-intervals}}

The preceding sections have characterized the structure of citation
variability across repeated samples. We now turn to the core inferential
question: given a single sample of responses, how can a practitioner
quantify the uncertainty in the resulting visibility estimates?

Bootstrap resampling with replacement provides a tractable solution. For
a single sample of \emph{N} responses, we generate B bootstrap
replicates by sampling responses with replacement and recomputing the
visibility metric for each replicate. The resulting distribution of
bootstrap estimates provides an empirical approximation of the sampling
distribution of the metric, from which confidence intervals (CI) can be
computed. Because citation share is a ratio (domain citations / total
citations), both numerator and denominator are recomputed from each
bootstrap replicate; resampling at the response level correctly
propagates the dependence structure.

\begin{figure}[htbp]
\centering
\includegraphics[width=\textwidth,height=0.9\textheight,keepaspectratio]{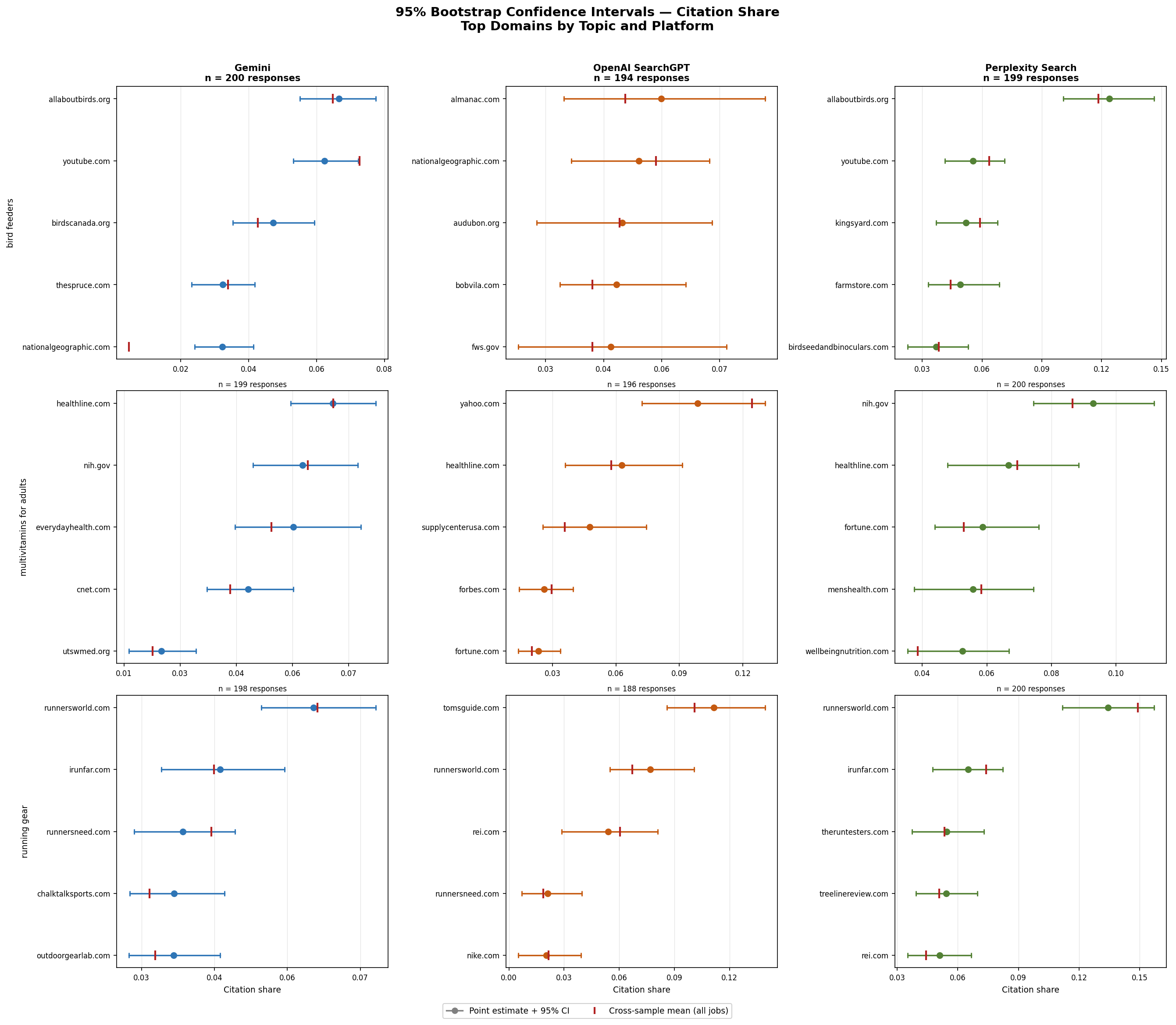}
\caption{95\% bootstrap confidence intervals for citation share. $3\times3$ grid: rows = topics, columns = platforms. For each panel, the dots represent the point estimates from the baseline sample, the error bars show the 95\% confidence intervals (1,000 bootstrap replicates), and the red tick marks show the cross-sample mean across all nine daily jobs.}
\label{fig:bootstrap}
\end{figure}

Figure~\ref{fig:bootstrap} presents 95\% bootstrap confidence intervals for each topic on
each platform.

At a 95\% confidence level, the span of the confidence intervals is
wide. For most of the frequently cited domains on SearchGPT, the span is
from 3 to 6 percentage points. For Gemini and Perplexity, the intervals
are narrower but still consequential. The top-ranked domains have the
largest citation shares and the largest CI spans:
\emph{everydayhealth.com} (multivitamins on Gemini) has a citation share
of 6.0\% and a CI span of 3.2 percentage points, and
\emph{runnersworld.com} (running gear on Perplexity) a citation share of
13.4\% and a CI span of 4.7 percentage points. Given the span of these
confidence intervals, apparent differences between many of the point
estimates---for any topic on any platform---are statistically
indistinguishable from sampling noise, as suggested by the many
overlapping confidence intervals.

It's worth noting that the bootstrap CI computed from a single sample
will not always contain the cross-sample mean. For example, the point
estimate for \emph{nationalgeographic.com} (bird feeders on Gemini) is
0.032, which lies far above the cross-sample mean of 0.005, and the CI
of 0.024-0.042 fails to capture the long-run value. A practitioner
relying on this single sample would rank \emph{nationalgeographic.com}
as a top-cited domain on Gemini---a conclusion that all subsequent
samples contradict. Such misses are not inherently pathological,
however: with a 95\% confidence interval we expect, by definition, that
approximately 1 in 20 estimates will fall outside the interval purely
due to sampling variability. However, the magnitude of the deviation in
cases like \emph{nationalgeographic.com} illustrates how a single
snapshot can still produce highly misleading conclusions about domain
visibility.

\hypertarget{effect-of-sample-size-on-confidence-interval-width}{%
\subsection{Effect of Sample Size on Confidence Interval
Width}\label{effect-of-sample-size-on-confidence-interval-width}}

\begin{figure}[htbp]
\centering
\includegraphics[width=\textwidth,height=0.9\textheight,keepaspectratio]{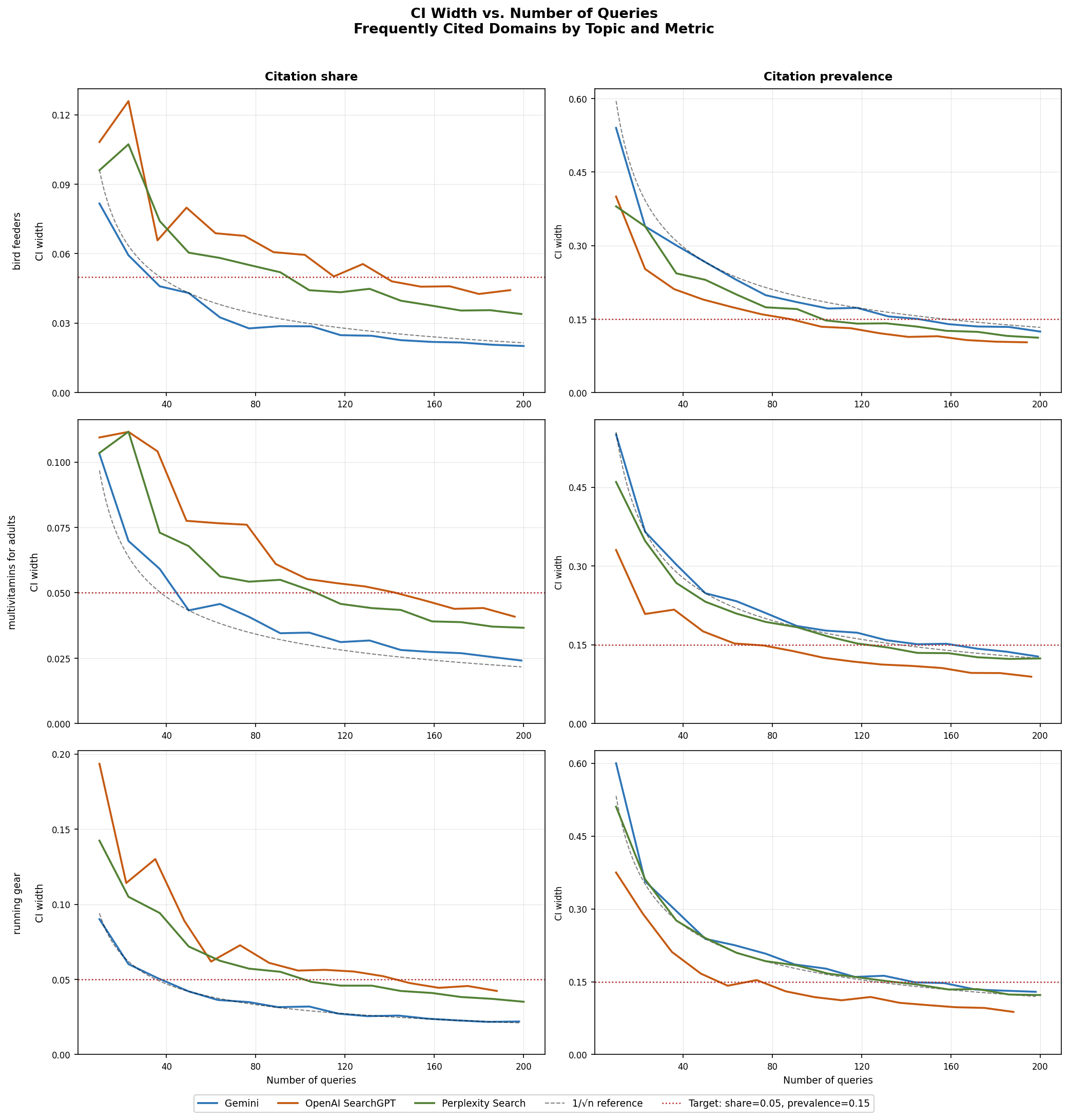}
\caption{CI width as a function of number of queries. $3\times2$ grid: rows = topics (bird feeders, multivitamins for adults, running gear), left column = citation share, right column = citation prevalence. One line per platform (Gemini: blue; SearchGPT: orange; Perplexity: green). Gray dashed curve: $1/\sqrt{n}$ reference. Red dotted horizontal line: target width (share: 0.05; prevalence: 0.15).}
\label{fig:ciwidth}
\end{figure}

Figure~\ref{fig:ciwidth} plots CI width as a function of the number of queries for each
platform-topic combination, separately for citation share (left column)
and citation prevalence (right column). Each line represents the maximum
CI width across all frequently cited domains for a given platform and
topic. This reflects a worst-case precision bound: if even one domain
has a CI too wide to interpret, the sample cannot reliably distinguish
among domain pairs.

The dominant domain typically drives this maximum. Citation shares in
the range 0.05-0.15 lie near the middle of the proportion scale, where
variance \emph{p}(1$-$\emph{p}) is largest in absolute terms. A domain
with a 12\% share therefore carries substantially more absolute variance
in its citation counts than a domain at 1\%, even though it may appear
more stable in relative terms.

Each panel overlays a $1/\sqrt{n}$ reference curve representing the CI
width expected under ideal conditions: independent, identically
distributed responses drawn from a fixed underlying distribution. For a
95\% confidence interval on a proportion p, the theoretical width is
$3.92\cdot\sqrt{p(1-p)/n}$, where 3.92 = $2\times1.96$ spans both
tails of the normal distribution. The reference curves for citation
share and citation prevalence are anchored at different vertical scales
because prevalence values (0.15-0.45) live closer to 0.5, where
\emph{p}(1$-$\emph{p}) is larger, while citation share values (0.03-0.15)
lie in a region where \emph{p}(1$-$\emph{p}) is smaller. Both curves have
the same $1/\sqrt{n}$ shape; only the scaling constant differs.

We define target CI widths of 0.05 for citation share and 0.15 for
citation prevalence as practical precision benchmarks. Deviations from
the reference curve are informative in two ways. First, vertical
distance from the reference indicates how efficiently additional queries
contribute to precision. Curves consistently above the reference suggest
that responses are not fully independent. Positive correlation across
queries or drift in the citation distribution reduces the effective
information contributed by each query. Curves below the reference
typically reflect unusually concentrated citation distributions. Second,
local bumps, where CI width temporarily increases or stalls before
resuming its decline, indicate within-sample non-stationarity: the
citation distribution shifts during part of the sample before
stabilizing again.

For citation share, Gemini converges fastest across all topics. The
Gemini CI width crosses the 0.05 target at approximately \emph{n} $\approx$ 30
for bird feeders and \emph{n} $\approx$ 40-50 for multivitamins and running
gear. The decline is smooth and closely tracks the $1/\sqrt{n}$ reference
curve, indicating that Gemini's citation distribution is relatively
stable within the sample and that queries contribute approximately the
expected amount of statistical information.

Perplexity represents the intermediate case. CI widths cross the target
at roughly \emph{n} $\approx$ 90-100 across topics. Convergence is less smooth
than Gemini's, with visible bumps in the bird feeders and multivitamins
curves indicating localized shifts in the citation distribution, but all
three topics reach the target within the 200-query sample.

SearchGPT shows the slowest and least monotonic convergence. The bumps
visible in all three citation-share curves indicate that the underlying
distribution shifts across queries. Although the curves begin to smooth
after \emph{n} $\approx$ 120, this should not be interpreted as evidence that
collecting 120 or more queries resolves the instability. As \emph{n}
approaches the full sample size of 200, subsamples are drawn from an
increasingly exhausted pool, forcing the bootstrap variance to
stabilize. Consequently, the curve must flatten as n$\rightarrow$200, regardless of
the true volatility of the system. Whether the underlying distribution
is stable across repeated collections is therefore a separate question,
addressed in Sections 5.4 and 5.5, where SearchGPT's log-std values and
daily time series show persistent citation volatility across all nine
samples.

For citation prevalence, the platform ordering reverses. SearchGPT
converges fastest, crossing the 0.15 target earliest in every topic
(\emph{n} $\approx$ 60-80), followed by Perplexity (\emph{n} $\approx$ 100-140), and
then Gemini (\emph{n} $\approx$ 140-150). This reversal reflects structural
differences between the two metrics: citation share is driven by the
absolute volume of citations to dominant domains, whereas prevalence
measures how broadly a domain appears across responses, which is easier
to estimate for platforms that produce fewer citations per response.

These convergence curves provide practical guidance for minimum sample
sizes. For citation share, Gemini requires roughly \emph{n} $\approx$ 40-50
queries to achieve 95\% confidence intervals spanning five percentage
points, Perplexity requires about \emph{n} $\approx$ 100, and SearchGPT requires
\emph{n} $\ge$ 150, reflecting the increasing degree of non-stationarity in
their citation distributions. For citation prevalence, SearchGPT
requires \emph{n} $\approx$ 60-80 queries to achieve intervals spanning fifteen
percentage points, whereas Gemini and Perplexity may require nearly
twice as many (\emph{n} $\approx$ 140-150).

Collecting fewer queries than these thresholds results in proportionally
wider confidence intervals, making it difficult to distinguish signal
from noise. For example, a domain with a true citation share of 8\% and
a CI spanning ten percentage points cannot be meaningfully distinguished
from domains at 3\% or 13\%; apparent differences or changes remain
statistically uninterpretable.

Non-monotonic convergence further complicates interpretation. If CI
width declined smoothly with \emph{n}, practitioners could apply an
early-stopping rule: collect queries until the CI width falls below the
target and then stop. The SearchGPT curves show why this approach is
unsafe. CI width can temporarily narrow and then widen again as
additional queries are added, reflecting shifts in the citation
distribution. A practitioner who stopped at a fortuitously narrow
intermediate point would report a CI that understates the true
uncertainty. The safer approach is therefore to commit in advance to a
fixed sample size based on prior measurements of the platform and topic,
rather than using the running CI width as a stopping criterion.

The convergence curves presented here are a foundation for a more
practical question that this paper deliberately leaves open: given a
target CI width and a platform, what is the minimum number of queries a
practitioner should collect? The platform- and topic-specific crossing
points identified in this section---and the non-convergence of SearchGPT
for bird feeders citation share---provide the empirical basis for that
guidance. Deriving it rigorously, as a function of citation distribution
parameters, is reserved for follow-up work and identified as a priority
in Section 10.

\hypertarget{distribution-wide-rank-stability}{%
\subsection{Distribution-Wide Rank
Stability}\label{distribution-wide-rank-stability}}

The preceding sections quantify measurement uncertainty primarily for
the top-5 frequently-cited domains. This is a natural starting point,
these are the domains a practitioner is most likely to monitor, but it
raises a question about generalizability: is the instability we observe
concentrated in the head of the distribution, where the confidence
intervals are widest, or does it extend across the full ranking? If rank
orderings are unstable throughout the frequently-cited domain set, then
the measurement uncertainty problem documented in Sections 5.6 and 5.7
is not a feature of just the top-ranked domains but a property of the
entire distribution. That would in turn mean that the value of
confidence intervals extends equally to the entire distribution, not
just to the top-ranked domains that happen to receive focused attention.

The weighting is a deliberate methodological choice motivated by the
noise properties of low-share domains. In an unweighted Spearman, a
domain moving from one citation to two citations can jump many rank
positions among the hundreds of low-share domains in the
frequently-cited set. Because such rank changes are numerous and driven
by single-citation fluctuations rather than structural shifts, an
unweighted correlation has high bootstrap variance: the CI for each pair
becomes very wide, and many pairs are classified as insufficient (CI
width \textgreater{} 0.25) even when the rank ordering of the
commercially relevant head of the distribution is highly stable.
Weighting by citation share concentrates the correlation on rank changes
that are substantively meaningful, damping the noise from the low-share
tail without discarding it entirely. The result is a measure that speaks
to distribution-wide stability in a way that is both statistically
tractable and interpretively appropriate: a rank swap between the 1st
and 5th domain is correctly treated as more consequential than a rank
swap between the 200th and 201st.

Confidence intervals for each pair's \emph{$\rho$} are constructed by
bootstrapping over domains: the set of frequently-cited domains is
resampled with replacement, and the weighted Spearman correlation is
recomputed on each resample. The 95\% CI is the interval between the
2.5th and 97.5th percentiles of the resulting bootstrap distribution
across 1000 iterations. This is a standard percentile bootstrap, without
bias correction or Fisher \emph{z}-transformation.

A consequence of this construction is that the point estimate can
occasionally fall outside the CI band. Two properties interact to
produce this. First, the bootstrap distribution of weighted \emph{$\rho$} is
skewed when the point estimate is close to 1: because \emph{$\rho$} is
bounded above by 1, bootstrap replicates are constrained to produce
correlations no greater than 1, compressing the upper tail and pulling
the lower percentile upward. Second, resampling domains with replacement
redistributes the weights (e.g., a domain drawn twice receives double
weight) changing the effective contribution of each domain to the
correlation relative to the point estimate computed on the full,
naturally-weighted set. The combination of boundary compression and
weight redistribution can produce a bootstrap distribution whose lower
percentile exceeds the observed point estimate. This is a known
limitation of the percentile bootstrap near parameter boundaries and
does not indicate a computational error. It arises precisely in the
cases where stability is highest, and is therefore conservative in the
direction of the paper's argument.

\begin{figure}[htbp]
\centering
\includegraphics[width=\textwidth,height=0.9\textheight,keepaspectratio]{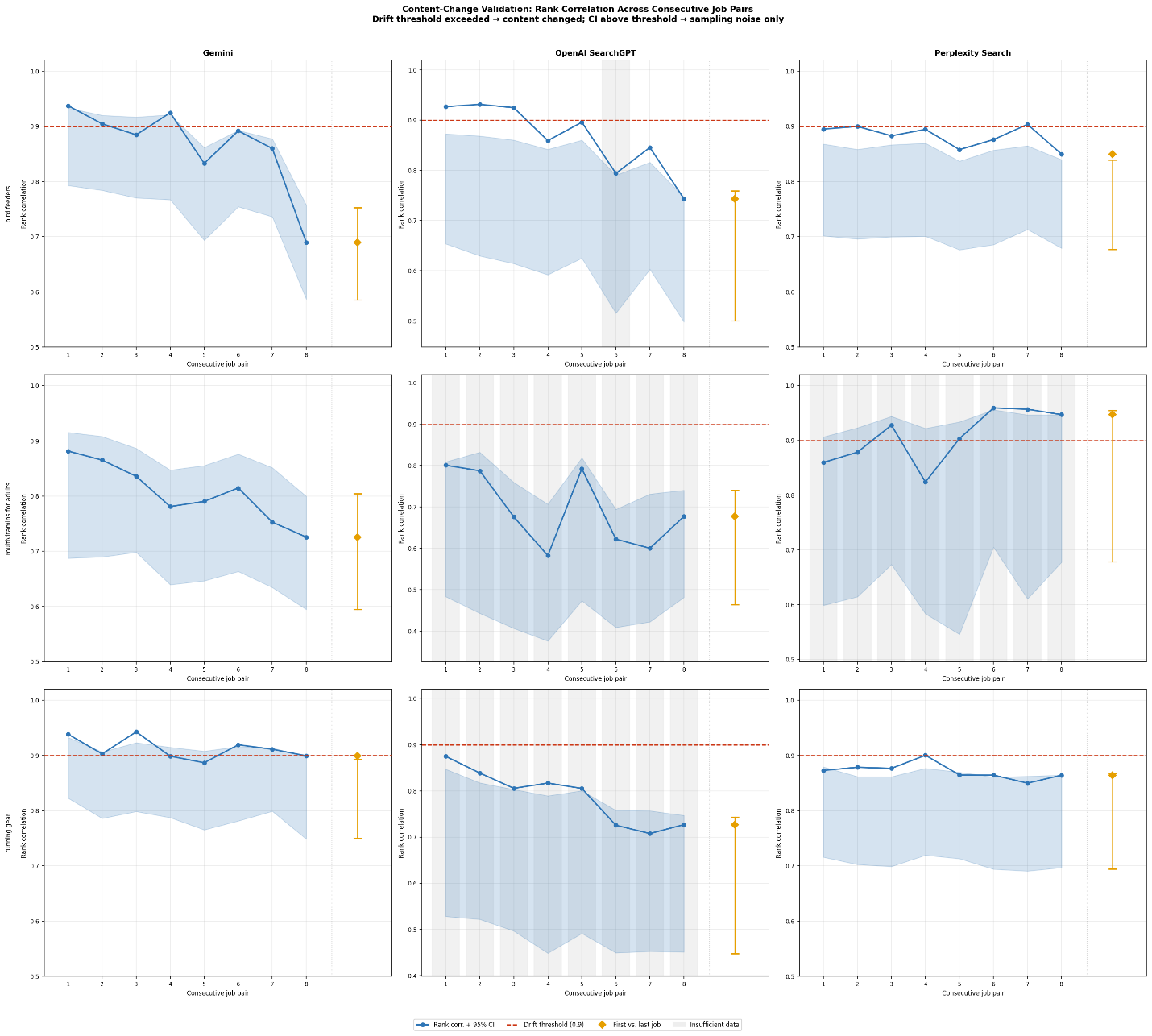}
\caption{Distribution-wide rank stability: rank correlation across consecutive job pairs. $3\times3$ grid: rows = topics (bird feeders, multivitamins for adults, running gear), columns = platforms (Gemini, SearchGPT, Perplexity). Blue line with 95\% CI band: Spearman rank correlation for each consecutive job pair. Red dashed line: stability threshold ($\rho = 0.9$). Gray shading: insufficient data pairs (CI width $>0.25$). Yellow diamond: span comparison (first vs.\ last job).}
\label{fig:rankstab}
\end{figure}

Figure~\ref{fig:rankstab} plots rank correlation across consecutive job pairs for all
nine platform-topic combinations, with the 0.9 threshold shown as a red
dashed line, gray shading for insufficient pairs (where the CI width
exceeds 0.25), and a yellow diamond for the span comparison (first job
vs. last job). Table 6 summarizes the results numerically.

\begin{table}[htbp]
\centering
\caption{Distribution-wide rank stability summary, one row per platform-topic combination. Pairs = total consecutive job pairs; Sufficient = pairs with bootstrap CI width $\le 0.25$; Stable = sufficient pairs with Spearman $\rho \ge 0.9$; Mean Rank $\rho$ and Mean CI Width are averaged over sufficient pairs; Span Rank $\rho$ compares the first and last job; Span Drift indicates whether a statistically and practically significant drift was detected over the nine-day window.}
\label{tab:rankstab}
\resizebox{\textwidth}{!}{%
\begin{tabular}{llccccccc}
\toprule
Platform & Topic & Pairs & Sufficient & Stable & Mean Rank $\rho$ & Mean CI Width & Span Rank $\rho$ & Span Drift \\
\midrule
Gemini & bird feeders & 8 & 8 & 4 & 0.866 & 0.150 & 0.689 & TRUE \\
Gemini & multivitamins & 8 & 8 & 2 & 0.806 & 0.209 & 0.726 & TRUE \\
Gemini & running gear & 8 & 8 & 7 & 0.913 & 0.127 & 0.900 & TRUE \\
\addlinespace
SearchGPT & bird feeders & 8 & 5 & 0 & 0.905 & 0.225 & 0.744 & FALSE \\
SearchGPT & multivitamins & 8 & 0 & 0 & --- & --- & 0.677 & FALSE \\
SearchGPT & running gear & 8 & 0 & 0 & --- & --- & 0.727 & FALSE \\
\addlinespace
Perplexity & bird feeders & 8 & 8 & 0 & 0.882 & 0.163 & 0.850 & TRUE \\
Perplexity & multivitamins & 8 & 1 & 1 & 0.959 & 0.250 & 0.947 & FALSE \\
Perplexity & running gear & 8 & 8 & 0 & 0.872 & 0.163 & 0.864 & TRUE \\
\bottomrule
\end{tabular}}
\end{table}

The results differ sharply by platform and topic. Gemini running gear is
the cleanest stability result in the dataset: all eight pairs are
sufficient, seven are stable, mean \emph{$\rho$} = 0.913, and the span
comparison \emph{$\rho$} = 0.900---right at the threshold. Gemini bird
feeders and multivitamins tell a different story: all eight pairs are
sufficient in both cases, but only four and two pairs, respectively, are
stable, mean \emph{$\rho$} falls to 0.866 and 0.806, and span \emph{$\rho$} drops
to 0.689 and 0.726---well below the threshold. Span drift is detected
for all three Gemini topics, indicating that cumulative drift over the
nine-day window is a platform-level property for Gemini regardless of
topic.

SearchGPT presents the most extreme picture. Bird feeders has five
sufficient pairs but zero stable ones, with a mean CI width of
0.225---close to the sufficiency boundary. Multivitamins and running
gear have zero sufficient pairs entirely, making mean \emph{$\rho$} and mean
CI width undefined; span \emph{$\rho$} values of 0.677 and 0.727 can be
computed but carry no inferential weight given the insufficient pair
counts. No span drift is detected for SearchGPT, but this is an artefact
of the insufficient data rather than evidence of stability---there is
simply not enough precision to detect drift that may be present.

Perplexity occupies a middle position. Bird feeders and running gear
each have all eight pairs sufficient and span \emph{$\rho$} values of 0.850
and 0.864, with span drift detected in both. Multivitamins has only one
sufficient pair---the mean CI width of 0.250 confirms it sits at the
boundary of the sufficiency criterion---but that one pair is stable and
span \emph{$\rho$} = 0.947. The wide CI bands for Perplexity reflect
head-domain concentration rather than instability per se: a distribution
dominated by one or two high-share domains has fewer effective rank
positions to correlate, inflating CI width.

An important cross-cutting finding: short-window consecutive-pair
stability does not imply long-window stability. For Gemini bird feeders
and multivitamins, each consecutive pair has \emph{$\rho$} \textgreater{}
0.80 with CIs straddling the threshold, yet the span comparison falls to
approximately 0.69-0.73---a cumulative drift that would be invisible
from any single adjacent-pair comparison. Long-window span analysis is
therefore a necessary complement to consecutive-pair monitoring.

The rank stability results confirm that distribution-wide instability is
the norm rather than the exception across platforms and topics. For the
combinations where rank correlation falls below the sufficiency
threshold, Gemini bird feeders and multivitamins across the nine-day
window, and SearchGPT across most combinations, uncertainty is not
confined to individual domain shares at the head of the distribution but
propagates through the entire ranking. Confidence intervals are
therefore not a refinement useful only for comparing the top two or
three domains: they are a necessary tool for any inference drawn from a
citation distribution, at any rank position.

The sufficiency criterion (\emph{$\rho$} $\ge$ 0.9) used here is a practical
benchmark rather than a formally derived threshold. Whether this
criterion is appropriate, and how it should be calibrated as a function
of the number of frequently-cited domains and the intended use of the
ranking, is an open question. Establishing principled sufficiency
thresholds and connecting them to minimum sample size guidance is a
priority for future work identified in Section 10.

\hypertarget{content-change-validation}{%
\section{Content-Change Validation}\label{content-change-validation}}

The findings in Section 5 document substantial citation variability
across platforms, topics, and sampling occasions. A natural alternative
explanation is that the content of cited pages changed between samples,
making some pages more or less relevant and thereby altering citation
probabilities. This section reports a methodological control designed to
rule out that explanation.

\hypertarget{approach}{%
\subsection{Approach}\label{approach}}

Using the content-change monitoring infrastructure described in Section
4.5, we identify domains for which citation counts changed significantly
between adjacent daily samples. For each flagged domain, we compare the
SHA-256 checksums of human-readable page content across the
corresponding job pairs. A stable checksum indicates that the page
content did not change; a changed checksum indicates a potential content
update.

The drift detection procedure uses the first daily sample as a frozen
baseline and compares each subsequent sample against it using a
chi-squared test with both statistical and practical significance
thresholds. A domain is flagged if both (a) the chi-squared p-value
falls below 0.05 and (b) the total share delta exceeds the practical
significance threshold.

Content status for each domain-job transition is classified as follows.
A transition is classified as unchanged if the content hash is identical
to the previous job. It is classified as changed if at least one URL for
the domain has a different hash from the previous job. It is classified
as unknown if hash data are unavailable for this or the preceding job.
By construction, the first job in any panel receives unknown status,
since there is no prior job to compare against.

\hypertarget{results}{%
\subsection{Results}\label{results}}

\begin{figure}[htbp]
\centering
\includegraphics[width=\textwidth,height=0.9\textheight,keepaspectratio]{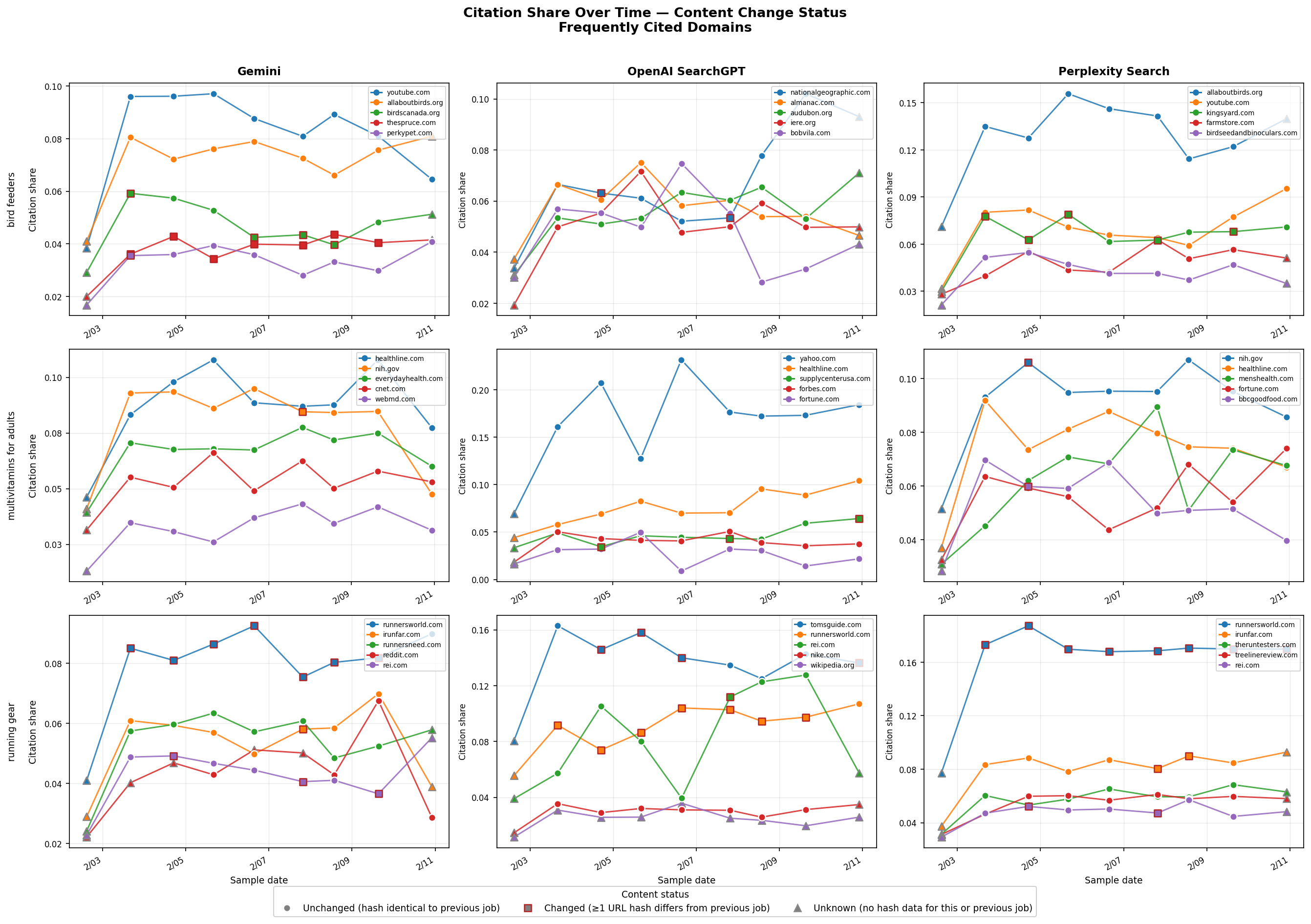}
\caption{Citation share over time---content-change status. $3\times3$ grid: rows = topics, columns = platforms. Top-5 frequently-cited domains per panel. Daily samples. Marker shapes encode content status: circle = unchanged (hash identical to previous job); square = changed ($\ge$1 URL hash differs); triangle = unknown (no hash data for this or previous job). Triangles are confined to the first job in every panel.}
\label{fig:contentchange}
\end{figure}

Figure~\ref{fig:contentchange} plots citation share over time for the top-5 frequently-cited
domains per panel, with each data point shaped to indicate the content
status of that domain at that job relative to the previous job.

Triangle markers (unknown) are confined to the first job in every panel,
confirming that hash coverage is complete from the second job onward.
The overwhelming majority of data points are circles, indicating that
the underlying pages were stable across consecutive job transitions for
most domains throughout the nine-day window. The citation share
volatility documented in Sections 5.4, 5.5, and 5.7 therefore cannot be
attributed to changes in cited page content: the oscillations, rank
crossings, and platform-level differences occur against a background of
largely stable source material.

A subset of domains display square markers on every job transition after
the baseline. The most prominent example is \emph{runnersworld.com} on
Perplexity (running gear topic), which shows a single unknown triangle
on the first job followed by changed squares on all subsequent jobs.
This pattern is consistent with server-side rendering variation: access
timestamps, session tokens, advertisement slots, or similar ephemeral
page elements that produce a different byte sequence on each HTTP
request without altering human-readable content. Hash-based change
detection is a conservative signal: identical hashes definitively rule
out content change, but differing hashes are a necessary, not
sufficient, condition for meaningful editorial change. We report these
cases transparently and do not attempt to distinguish rendering
variation from genuine content change.

The substantive finding is unaffected by these ambiguous cases. Even for
domains where hash-based stability cannot be confirmed, citation share
volatility of the magnitude observed here (rank crossings within a
single four-hour high-frequency window, CI widths of 3-7 percentage
points on daily data) cannot plausibly be explained by incremental
server-side changes that produce hash churn. The variability is
structural, not content-driven.

\hypertarget{discussion}{%
\section{Discussion}\label{discussion}}

\hypertarget{implications-for-visibility-measurement}{%
\subsection{Implications for Visibility
Measurement}\label{implications-for-visibility-measurement}}

The central implication is that single-run measurements of domain
visibility in generative search cannot be interpreted as reliable
estimates of an underlying true citation rate. The confidence intervals
reported in Section 5.6 show that observed values for a given domain can
vary substantially across repeated samples, and that many apparent
differences between domains are not statistically distinguishable. This
has concrete consequences for how practitioners should interpret
visibility measurements. Ranking changes between samples, apparent
surges or declines in citation share, and comparisons between domains
that are close in the distribution may all be artifacts of sampling
noise rather than signals of genuine change.

The appropriate inferential unit is not the point estimate but the
confidence interval. A domain with citation share 6\% $\pm$ 2 percentage
points (95\% CI: 4\%-8\%) is in a meaningfully different measurement
state than one reported as simply ``6\%.'' The former acknowledges the
uncertainty inherent in the measurement; the latter conveys false
precision. Bootstrap confidence intervals, computed at the response
level using the procedure described in Section 5.6, provide this
information with minimal additional computational cost.

The rank stability analysis in Section 5.8 extends this implication
beyond the top-ranked domains. A practitioner might reasonably suppose
that measurement uncertainty is primarily a problem for closely-ranked
domains near the head of the distribution, and that lower-ranked domains
can be safely ignored or that their relative ordering is stable. The
distribution-wide rank correlation results contradict this: for most
platform-topic combinations, the rank ordering of the full
frequently-cited domain set is either unstable across samples or cannot
be assessed with sufficient precision. Confidence intervals are
therefore not a refinement applicable only to the top two or three
domains---they are a necessary tool for any inference drawn from a
citation distribution at any rank position.

\hypertarget{implications-for-visibility-optimization-research}{%
\subsection{Implications for Visibility Optimization
Research}\label{implications-for-visibility-optimization-research}}

Work in the GEO tradition (Aggarwal et al.) reports visibility
improvements from content interventions without confidence intervals.
The findings of this paper suggest that such claims require statistical
validation: an observed improvement in citation share must be shown to
exceed the noise floor of the measurement process before it can be
attributed to the intervention. Bootstrap confidence intervals on both
the baseline and post-intervention measurements provide the necessary
inferential framework.

This requirement is not merely theoretical. Given the CI widths
documented in this paper, an intervention that improves a domain's
SearchGPT citation share from 8\% to 11\% cannot be attributed to the
intervention with confidence: the improvement of 3 percentage points
falls within the typical CI width for a SearchGPT domain. Only larger
effects, or effects validated by repeated pre/post measurement, are
distinguishable from noise.

\hypertarget{platform-differences}{%
\subsection{Platform Differences}\label{platform-differences}}

The three platforms studied here differ in several empirically
documented ways that affect both measurement design and interpretation.
Citation volume differs by roughly six times across platforms: Gemini
produces approximately 40-43 citations per response, Perplexity
approximately 20-22, and SearchGPT approximately 6-7. This drives
Gemini's larger frequently-cited domain set (142-157 per topic) and
longer power-law tail relative to SearchGPT (84-95) and Perplexity
(95-114).

Rank stability differs: Perplexity's top-cited domains have the lowest
domain-level log-std in the dataset (\emph{runnersworld.com}: 0.062;
\emph{nih.gov}: 0.072; \emph{allaboutbirds.org}: 0.095), while Gemini's
tail frequently-cited domains include extreme outliers
(\emph{gardeningknowhow.com}: log-std = 1.93). Response consistency
differs: SearchGPT shows a bimodal Jaccard distribution (either
near-perfect repeat or complete divergence) and nine domains with
log-std = 0.0 (deterministic citation behavior for those domains), while
Gemini and Perplexity show unimodal distributions centered around
Jaccard 0.30 and 0.50, respectively.

These differences have direct implications for measurement design. The
number of queries needed to achieve a given CI width is platform- and
topic-specific. The interpretation of single-sample estimates carries
different risks depending on whether the platform's citation behavior is
primarily deterministic, stochastic, or intermediate. A measurement
protocol designed for Gemini (where \emph{n} = 40-50 might be
sufficient) is inadequate for SearchGPT, where no fixed-\emph{n}
protocol achieves the target precision within a realistic budget. For
SearchGPT, the problem is not only that more queries are needed but that
within-sample non-stationarity---evidenced by the non-monotonic
convergence curves in Section 5.7---means the citation distribution
itself shifts across the query sequence. This makes early-stopping rules
unreliable and requires committing to a fixed \emph{n} based on prior
knowledge of the platform. Rank stability compounds the picture:
SearchGPT multivitamins and running gear have zero sufficient pairs in
the rank correlation analysis, meaning the distribution-wide rank
ordering cannot be assessed at all with the current sampling budget.

\hypertarget{limitations}{%
\section{Limitations}\label{limitations}}

We identify the following limitations of this study.

\begin{itemize}
\item
  Three topics. The study covers three consumer product topics chosen to
  represent varied market structures. Generalization to other domains,
  including B2B topics, navigational queries, and rapidly evolving news
  topics, requires further study.
\item
  LLM-generated queries. Queries were generated by ChatGPT, which may
  have implicit associations between topics and sources that shape the
  query distribution. This is a departure from using observed user query
  logs and may affect the representativeness of the query set.
\item
  Gemini exclusion from high-frequency regime. Google Gemini was
  excluded from the ten-minute interval experiment due to API rate
  limits. The high-frequency findings therefore cover only Perplexity
  Search and OpenAI SearchGPT.
\item
  Scraping coverage. Content checksums were computed from HTML scraped
  during data collection. Coverage is incomplete due to server-side
  blocking (Cloudflare, CAPTCHA) and format limitations (video content,
  PDFs). Conclusions about content stability are based on the subset of
  URLs for which scraping succeeded.
\item
  Single observation window. All data were collected over a period of
  approximately nine days. Longer-term dynamics, including seasonal
  variation and major platform updates, are not captured.
\item
  Infrequently cited domains. The statistical treatment of zero-inflated
  citation data is out of scope for this paper and represents an open
  methodological challenge.
\end{itemize}

\hypertarget{conclusion}{%
\section{Conclusion}\label{conclusion}}

Generative search engines are non-deterministic systems, and the
citation visibility metrics derived from their outputs are consequently
random variables subject to both system-level stochasticity and
measurement uncertainty. This paper has demonstrated empirically that
this variability is large enough to materially affect analytical
conclusions: many apparent differences between domains, and many
apparent changes across samples, are not statistically distinguishable
from noise.

Bootstrap confidence intervals provide a principled and tractable method
for quantifying this uncertainty. Applied to citation count, share, and
prevalence metrics, they reveal that single-run measurements carry
substantial imprecision that is invisible when results are reported as
point estimates. The intervals are wide enough to render common
analytical claims (e.g., domain A outranks domain B; visibility improved
after intervention X) untestable without repeated sampling.

The content-change validation confirms that observed variability is
predominantly structural: the underlying source pages are largely
stable, and the citation oscillations reflect engine behavior rather
than changing content. The power-law characterization of citation
distributions provides a theoretical grounding for log-space dispersion
analysis and motivates the distinction between frequently and
infrequently cited domains that is central to the measurement framework.
The distribution-wide rank stability analysis confirms that this
measurement uncertainty is not confined to the top-ranked domains: rank
instability extends across the full frequently-cited domain set for most
platform-topic combinations, establishing that confidence intervals are
necessary for inference at any rank position, not merely among the most
visible domains.

We hope these findings motivate a shift in how visibility measurement is
practiced and reported in the growing literature on generative search
optimization toward a framework that treats citation visibility as what
it is: a probabilistic quantity estimated from a finite, noisy sample.

\hypertarget{future-work}{%
\section{Future Work}\label{future-work}}

The following directions are identified as natural continuations of this
work.

\begin{itemize}
\item
  Minimum sample size guidance. A follow-up study will provide
  principled guidance on how many queries are needed to achieve a target
  confidence interval width for citation share and prevalence metrics,
  as a function of citation count distribution parameters and the
  platform being measured.
\item
  Statistical treatment of infrequently cited domains. Zero-inflated
  count models and occupancy models from ecology offer potential
  frameworks for estimating citation probabilities for domains that
  appear in only some samples. Whether platform-level experimentation
  with grounding sources contributes to the observed presence/absence
  pattern is an empirical question for future study.
\item
  Longer observation windows. A study spanning weeks or months would
  enable characterization of longer-term dynamics, including seasonal
  patterns and the effects of platform architecture updates.
\item
  Additional topics and query types. Extending the analysis to B2B,
  navigational, and news-oriented topics would test the generalizability
  of the findings.
\item
  Real user queries. Replacing LLM-generated queries with observed user
  query logs would eliminate the query generation confound and improve
  ecological validity.
\end{itemize}

\bibliographystyle{plain}
\bibliography{references}

@inproceedings{aggarwal2024,
  author    = {Aggarwal, Pranjal and Murahari, Vishvak and Rajpurohit, Tanmay and Kalyan, Ashwin and Narasimhan, Karthik and Deshpande, Ameet},
  title     = {{GEO}: Generative Engine Optimization},
  booktitle = {Proceedings of the 30th ACM SIGKDD Conference on Knowledge Discovery and Data Mining (KDD '24)},
  pages     = {5--16},
  year      = {2024},
  publisher = {ACM},
  note      = {arXiv:2311.09735}
}

@inproceedings{bink2022,
  author    = {Bink, Markus and Zimmerman, Steven and Elsweiler, David},
  title     = {Featured Snippets and their Influence on Users' Credibility Judgements},
  booktitle = {Proceedings of the 2022 Conference on Human Information Interaction and Retrieval (CHIIR '22)},
  pages     = {113--122},
  year      = {2022},
  publisher = {ACM},
  doi       = {10.1145/3498366.3505766}
}

@inproceedings{cutrell2007,
  author    = {Cutrell, Edward and Guan, Zhiwei},
  title     = {What Are You Looking For? An Eye-Tracking Study of Information Usage in Web Search},
  booktitle = {Proceedings of the SIGCHI Conference on Human Factors in Computing Systems (CHI '07)},
  pages     = {407--416},
  year      = {2007},
  publisher = {ACM},
  doi       = {10.1145/1240624.1240691}
}

@inproceedings{ding2025,
  author    = {Ding, Yifan and Facciani, Matthew and Joyce, Elizabeth and Poudel, Amrit and Bhattacharya, Sanmoy and Veeramani, Balaji and Aguinaga, Salvador and Weninger, Tim},
  title     = {Citations and Trust in {LLM} Generated Responses},
  booktitle = {Proceedings of the AAAI Conference on Artificial Intelligence},
  volume    = {39},
  number    = {22},
  pages     = {23787--23795},
  year      = {2025},
  doi       = {10.1609/aaai.v39i22.34550},
  note      = {arXiv:2501.01303}
}

@inproceedings{dupret2008,
  author    = {Dupret, Georges E. and Piwowarski, Beno{\^i}t},
  title     = {A User Browsing Model to Predict Search Engine Click Data from Past Observations},
  booktitle = {Proceedings of the 31st Annual International ACM SIGIR Conference on Research and Development in Information Retrieval (SIGIR '08)},
  pages     = {331--338},
  year      = {2008},
  publisher = {ACM},
  doi       = {10.1145/1390334.1390392}
}

@book{efron1993,
  author    = {Efron, Bradley and Tibshirani, Robert J.},
  title     = {An Introduction to the Bootstrap},
  publisher = {Chapman and Hall/CRC},
  address   = {New York},
  year      = {1993},
  isbn      = {978-0-412-04231-7}
}

@inproceedings{gao2023,
  author    = {Gao, Tianyu and Yen, Howard and Yu, Jiatong and Chen, Danqi},
  title     = {Enabling Large Language Models to Generate Text with Citations},
  booktitle = {Proceedings of the 2023 Conference on Empirical Methods in Natural Language Processing (EMNLP '23)},
  pages     = {6465--6488},
  year      = {2023},
  publisher = {Association for Computational Linguistics},
  note      = {arXiv:2305.14627}
}

@inproceedings{he2025,
  author    = {He, Jiqun and Liu, Jiqun},
  title     = {Not All Transparency Is Equal: Source Presentation Effects on Attention, Interaction, and Persuasion in Conversational Search},
  booktitle = {Proceedings of the 2026 Conference on Human Information Interaction and Retrieval (CHIIR '26)},
  year      = {2025},
  publisher = {ACM},
  note      = {arXiv:2512.12207}
}

@inproceedings{hu2025,
  author    = {Hu, Desheng and Baumann, Jakob and Urman, Aleksandra and Lichtenegger, Elazar and Forsberg, Robin and Hannak, Aniko and Wilson, Christo},
  title     = {Auditing Google's {AI} Overviews and Featured Snippets: A Case Study on Baby Care and Pregnancy},
  booktitle = {Proceedings of the International AAAI Conference on Web and Social Media (ICWSM '26)},
  year      = {2025},
  note      = {arXiv:2511.12920}
}

@inproceedings{joachims2005,
  author    = {Joachims, Thorsten and Granka, Laura A. and Pan, Bing and Hembrooke, Helene and Gay, Geri},
  title     = {Accurately Interpreting Clickthrough Data as Implicit Feedback},
  booktitle = {Proceedings of the 28th Annual International ACM SIGIR Conference on Research and Development in Information Retrieval (SIGIR '05)},
  pages     = {154--161},
  year      = {2005},
  publisher = {ACM}
}

@inproceedings{li2024,
  author    = {Li, Yifei and Yue, Xiang and Liao, Zeyi and Sun, Huan},
  title     = {{AttributionBench}: How Hard is Automatic Attribution Evaluation?},
  booktitle = {Findings of the Association for Computational Linguistics: ACL 2024},
  pages     = {14919--14935},
  year      = {2024},
  publisher = {Association for Computational Linguistics},
  note      = {arXiv:2402.15089}
}

@inproceedings{liu2023,
  author    = {Liu, Nelson F. and Zhang, Tianyi and Liang, Percy},
  title     = {Evaluating Verifiability in Generative Search Engines},
  booktitle = {Findings of the Association for Computational Linguistics: EMNLP 2023},
  pages     = {7001--7025},
  year      = {2023},
  publisher = {Association for Computational Linguistics},
  note      = {arXiv:2304.09848}
}

@article{madaan2024,
  author  = {Madaan, Lovish and Singh, Aaditya K. and Schaeffer, Rylan and Poulton, Andrew and Koyejo, Sanmi and Stenetorp, Pontus and Narang, Sharan and Hupkes, Dieuwke},
  title   = {Quantifying Variance in Evaluation Benchmarks},
  journal = {arXiv preprint},
  year    = {2024},
  note    = {arXiv:2406.10229}
}

@inproceedings{venkit2025,
  author    = {Venkit, Pranav Narayanan and Laban, Philippe and Zhou, Yilun and Mao, Yixin and Wu, Chien-Sheng},
  title     = {Search Engines in the {AI} Era: A Qualitative Understanding to the False Promise of Factual and Verifiable Source-Cited Responses in {LLM}-based Search},
  booktitle = {Proceedings of the 2025 ACM Conference on Fairness, Accountability, and Transparency (FAccT '25)},
  pages     = {1325--1340},
  year      = {2025},
  publisher = {ACM},
  doi       = {10.1145/3715275.3732089}
}

@article{xu2025,
  author  = {Xu, Yumo and Qi, Peng and Chen, Jifan and Liu, Kaixin and Han, Ruixin and Liu, Lujun and Min, Bonan and Castelli, Vittorio and Gupta, Arshit and Wang, Zhiguo},
  title   = {{CiteEval}: Principle-Driven Citation Evaluation for Source Attribution},
  journal = {arXiv preprint},
  year    = {2025},
  note    = {arXiv:2506.01829}
}

\appendix
\counterwithin{table}{section}

\section{Citation Format Examples}

The following tables illustrate how citations are implemented in
Perplexity and SearchGPT. Both examples are responses to the query,
``top rated multivitamins for adults by experts.''

\begin{table}[htbp]
\centering
\caption{Perplexity citation example. Response to the query ``top rated multivitamins for adults by experts.}
\label{tab:perpex}
\footnotesize
\begin{tabular}{@{}p{0.95\textwidth}@{}}
\toprule
Response \\
\midrule
Based on expert reviews for 2026, the \textbf{top-rated multivitamins for adults} vary by category and specific needs:\newline\newline \textbf{Overall Best Multivitamins.} \textbf{Performance Lab\textregistered{} NutriGenesis\textregistered{} Multi} is recommended as the best overall multivitamin supplement, offering 27 vitamins and minerals in gender-specific formulations for men and women with clean, plant-based delivery~[1].\newline\newline \textbf{Ritual Essential for Men} is rated as the best overall multivitamin specifically for men~[3], while \textbf{Inessa Advanced multivitamin} (\pounds29.99) earns top marks as a premium option with a 5/5 star rating~[2].\newline\newline \textbf{Top-Rated by Category.} \emph{Best for meeting daily recommended values:} Swallow daily multivitamin (5/5 stars)~[2]. \emph{Best one-a-day:} HealthSpan Gold A-Z multivitamin (5/5 stars) at 13p per day~[2]. \emph{Best for women:} Dr Vegan Women's ProMulti (5/5 stars)~[2]. \emph{Best all-rounder capsule:} Viridian High-5 (4.5/5 stars)~[2]. \emph{Best budget option:} Healthspan Multivitality Pro (4/5 stars) at 19p per day~[2]. For those over 50, \textbf{Centrum Silver Adults 50+} and \textbf{Thorne Multi 50+ Daily} are recommended by experts~[3][4].\newline\newline [1] \url{https://www.performancelab.com/blogs/multi/best-multivitamin}\newline [2] \url{https://www.bbcgoodfood.com/review/best-multivitamins}\newline [3] \url{https://www.menshealth.com/health/a19545998/best-multivitamins-for-men/}\newline [4] \url{https://www.cspi.org/article/multiple-choice-how-find-best-multivitamin} \\
\bottomrule
\end{tabular}
\end{table}

\begin{table}[htbp]
\centering
\caption{SearchGPT citation example. Response to the query ``top rated multivitamins for adults by experts.}
\label{tab:searchex}
\footnotesize
\begin{tabular}{@{}p{0.95\textwidth}@{}}
\toprule
Response \\
\midrule
Selecting the right multivitamin can be challenging due to the multitude of options available. Based on expert reviews and evaluations from 2026, here are some top-rated multivitamins for adults:\newline\newline \textbf{Thorne Basic Nutrients 2/Day.} A comprehensive multivitamin providing 21 essential vitamins and minerals, including all eight B vitamins and vitamins A, C, D, E, and K. Notably, it offers 250\% of the daily value for vitamin D. (health.yahoo.com~--- \url{https://health.yahoo.com/wellness/nutrition/vitamins-supplements/article/best-multivitamin-for-2025-202710445.html})\newline\newline \textbf{Nature Made Multivitamin For Him.} An affordable, one-a-day multivitamin tailored for men, containing 22 essential nutrients, including vitamin D, zinc, and selenium. USP-verified for purity and potency. (health.yahoo.com~--- \url{https://health.yahoo.com/wellness/nutrition/vitamins-supplements/article/best-multivitamin-men-211005605.html})\newline\newline \textbf{Perelel Women's Daily Trio.} A multivitamin designed specifically for women, offering a blend of essential nutrients to support overall health. (health.yahoo.com~--- \url{https://health.yahoo.com/wellness/nutrition/vitamins-supplements/article/best-multivitamin-for-2025-202710445.html})\newline\newline \textbf{Nordic Naturals Men's Multivitamin.} Tailored for men's health, this multivitamin includes saw palmetto and nettle extract to support prostate health, along with essential nutrients like vitamin D and B vitamins. Third-party tested. (health.yahoo.com)\newline\newline \textbf{Garden of Life Organics Women's Once Daily.} A USDA-certified organic multivitamin formulated for women, providing over 17 essential nutrients sourced from whole foods. (health.yahoo.com)\newline\newline When choosing a multivitamin, consider factors such as your specific health needs, dietary restrictions, and lifestyle. It is advisable to consult with a healthcare professional to determine the most suitable option for you. \\
\bottomrule
\end{tabular}
\end{table}

\section{Per-Sample Summary Statistics}

\begin{table}[htbp]
\centering
\small
\caption{Gemini per-sample summary statistics.}
\label{tab:gemsamp}
\begin{tabular}{llccccccc}
\toprule
\multicolumn{4}{c}{} & \multicolumn{5}{c}{Citations} \\
\cmidrule(lr){5-9}
Platform & Topic & Sample & Responses & Mean & Median & Std & Min & Max \\
\midrule
Gemini & bird feeders & 1 & 200 & 41.1 & 36.0 & 21.3 & 9 & 110 \\
 &  & 2 & 200 & 42.4 & 36.0 & 22.3 & 7 & 130 \\
 &  & 3 & 198 & 39.9 & 35.0 & 20.7 & 7 & 154 \\
 &  & 4 & 200 & 42.8 & 36.0 & 23.2 & 7 & 147 \\
 &  & 5 & 200 & 41.3 & 36.0 & 21.8 & 10 & 112 \\
 &  & 6 & 194 & 40.9 & 35.0 & 22.2 & 11 & 158 \\
 &  & 7 & 200 & 41.1 & 36.5 & 21.0 & 10 & 112 \\
 &  & 8 & 198 & 42.2 & 36.5 & 22.0 & 6 & 104 \\
 &  & 9 & 199 & 51.1 & 44.0 & 27.5 & 7 & 152 \\
\addlinespace
 & multivitamins for adults & 1 & 199 & 38.1 & 35.0 & 17.4 & 3 & 109 \\
 &  & 2 & 198 & 40.2 & 38.0 & 18.3 & 3 & 100 \\
 &  & 3 & 197 & 39.7 & 35.0 & 20.0 & 9 & 138 \\
 &  & 4 & 199 & 37.6 & 35.0 & 20.1 & 7 & 169 \\
 &  & 5 & 199 & 40.7 & 36.0 & 20.9 & 5 & 142 \\
 &  & 6 & 199 & 38.9 & 36.0 & 18.9 & 3 & 106 \\
 &  & 7 & 199 & 38.8 & 35.0 & 17.9 & 8 & 92 \\
 &  & 8 & 199 & 38.3 & 35.0 & 18.1 & 5 & 112 \\
 &  & 9 & 198 & 44.1 & 41.5 & 23.9 & 8 & 197 \\
\addlinespace
 & running gear & 1 & 198 & 44.4 & 41.0 & 22.0 & 6 & 136 \\
 &  & 2 & 200 & 42.4 & 39.0 & 21.3 & 6 & 132 \\
 &  & 3 & 200 & 43.4 & 39.5 & 21.3 & 13 & 134 \\
 &  & 4 & 199 & 44.0 & 41.0 & 21.4 & 13 & 148 \\
 &  & 5 & 199 & 42.1 & 38.0 & 19.7 & 8 & 117 \\
 &  & 6 & 199 & 42.0 & 40.0 & 19.4 & 1 & 94 \\
 &  & 7 & 197 & 41.9 & 39.0 & 18.5 & 5 & 108 \\
 &  & 8 & 199 & 40.6 & 39.0 & 18.0 & 3 & 100 \\
 &  & 9 & 200 & 46.8 & 42.0 & 26.0 & 10 & 202 \\
\bottomrule
\end{tabular}
\end{table}

\begin{table}[htbp]
\centering
\small
\caption{SearchGPT per-sample summary statistics.}
\label{tab:searchsamp}
\begin{tabular}{llccccccc}
\toprule
\multicolumn{4}{c}{} & \multicolumn{5}{c}{Citations} \\
\cmidrule(lr){5-9}
Platform & Topic & Sample & Responses & Mean & Median & Std & Min & Max \\
\midrule
SearchGPT & bird feeders & 1 & 194 & 7.1 & 7.0 & 2.6 & 1 & 15 \\
 &  & 2 & 195 & 7.1 & 6.0 & 2.8 & 1 & 15 \\
 &  & 3 & 196 & 7.1 & 7.0 & 2.6 & 1 & 16 \\
 &  & 4 & 198 & 7.0 & 6.0 & 2.6 & 1 & 16 \\
 &  & 5 & 197 & 7.0 & 6.0 & 2.5 & 1 & 14 \\
 &  & 6 & 197 & 7.1 & 7.0 & 2.5 & 1 & 14 \\
 &  & 7 & 197 & 6.8 & 6.0 & 2.8 & 1 & 16 \\
 &  & 8 & 198 & 7.2 & 6.0 & 2.7 & 1 & 17 \\
 &  & 9 & 194 & 7.3 & 7.0 & 2.9 & 1 & 18 \\
\addlinespace
 & multivitamins for adults & 1 & 196 & 5.7 & 5.0 & 1.9 & 1 & 11 \\
 &  & 2 & 197 & 5.9 & 5.0 & 2.2 & 1 & 13 \\
 &  & 3 & 193 & 5.9 & 5.0 & 2.1 & 1 & 15 \\
 &  & 4 & 196 & 5.8 & 5.0 & 1.9 & 1 & 11 \\
 &  & 5 & 192 & 5.9 & 5.0 & 2.2 & 1 & 15 \\
 &  & 6 & 189 & 6.0 & 5.0 & 2.1 & 1 & 13 \\
 &  & 7 & 195 & 5.9 & 5.0 & 1.8 & 1 & 12 \\
 &  & 8 & 194 & 5.9 & 5.0 & 2.0 & 1 & 12 \\
 &  & 9 & 193 & 6.0 & 5.0 & 2.0 & 1 & 12 \\
\addlinespace
 & running gear & 1 & 188 & 6.3 & 5.0 & 2.6 & 1 & 18 \\
 &  & 2 & 186 & 6.7 & 5.0 & 2.6 & 1 & 14 \\
 &  & 3 & 185 & 6.8 & 6.0 & 2.5 & 1 & 18 \\
 &  & 4 & 186 & 6.2 & 5.0 & 2.3 & 1 & 14 \\
 &  & 5 & 187 & 6.7 & 5.0 & 2.5 & 1 & 16 \\
 &  & 6 & 191 & 6.6 & 5.0 & 2.5 & 1 & 14 \\
 &  & 7 & 187 & 6.7 & 6.0 & 2.6 & 1 & 17 \\
 &  & 8 & 192 & 6.6 & 5.0 & 2.7 & 1 & 19 \\
 &  & 9 & 191 & 6.6 & 5.0 & 2.6 & 1 & 16 \\
\bottomrule
\end{tabular}
\end{table}

\begin{table}[htbp]
\centering
\small
\caption{Perplexity per-sample summary statistics.}
\label{tab:perpsamp}
\begin{tabular}{llccccccc}
\toprule
\multicolumn{4}{c}{} & \multicolumn{5}{c}{Citations} \\
\cmidrule(lr){5-9}
Platform & Topic & Sample & Responses & Mean & Median & Std & Min & Max \\
\midrule
Perplexity & bird feeders & 1 & 199 & 21.8 & 21.0 & 9.4 & 3 & 57 \\
 &  & 2 & 200 & 22.9 & 22.0 & 10.2 & 7 & 72 \\
 &  & 3 & 200 & 22.0 & 20.0 & 9.7 & 3 & 48 \\
 &  & 4 & 200 & 21.4 & 20.0 & 9.7 & 5 & 55 \\
 &  & 5 & 200 & 22.1 & 20.0 & 9.9 & 4 & 56 \\
 &  & 6 & 200 & 21.9 & 20.0 & 9.9 & 3 & 49 \\
 &  & 7 & 199 & 22.7 & 21.0 & 9.8 & 4 & 63 \\
 &  & 8 & 200 & 21.5 & 21.0 & 9.4 & 2 & 48 \\
 &  & 9 & 200 & 21.9 & 20.0 & 9.5 & 7 & 58 \\
\addlinespace
 & multivitamins for adults & 1 & 200 & 19.9 & 18.0 & 8.5 & 5 & 51 \\
 &  & 2 & 200 & 20.0 & 19.0 & 8.5 & 6 & 45 \\
 &  & 3 & 200 & 20.3 & 19.0 & 7.8 & 6 & 44 \\
 &  & 4 & 200 & 20.1 & 18.0 & 8.3 & 6 & 49 \\
 &  & 5 & 200 & 19.8 & 19.0 & 8.1 & 5 & 46 \\
 &  & 6 & 200 & 19.6 & 19.0 & 8.0 & 5 & 41 \\
 &  & 7 & 200 & 20.0 & 19.0 & 8.2 & 7 & 42 \\
 &  & 8 & 199 & 20.1 & 19.0 & 8.4 & 6 & 52 \\
 &  & 9 & 200 & 20.5 & 18.5 & 9.0 & 3 & 49 \\
\addlinespace
 & running gear & 1 & 200 & 21.9 & 21.0 & 8.4 & 5 & 53 \\
 &  & 2 & 200 & 22.5 & 22.0 & 8.2 & 5 & 44 \\
 &  & 3 & 200 & 22.9 & 22.0 & 8.6 & 7 & 56 \\
 &  & 4 & 199 & 22.3 & 22.0 & 9.3 & 4 & 50 \\
 &  & 5 & 200 & 22.0 & 22.0 & 8.7 & 4 & 48 \\
 &  & 6 & 199 & 22.8 & 22.0 & 9.6 & 4 & 47 \\
 &  & 7 & 200 & 22.8 & 22.0 & 8.7 & 7 & 48 \\
 &  & 8 & 200 & 22.9 & 23.0 & 8.4 & 5 & 48 \\
 &  & 9 & 200 & 22.4 & 22.0 & 8.8 & 6 & 49 \\
\bottomrule
\end{tabular}
\end{table}

\end{document}